\documentclass[a4paper]{article}
\usepackage{amsmath,amssymb,amsfonts,amsthm}
\usepackage[titletoc,title]{appendix}
\usepackage{color}
\numberwithin{equation}{section}

\newcommand{\bea}{\begin{eqnarray}}
\newcommand{\eea}{\end{eqnarray}}
\newcommand{\be}{\begin{equation}}
\newcommand{\ee}{\end{equation}}
\newcommand{\nn}{\nonumber \\}

\begin{document}
\begin{titlepage}

\vfill \vfill \vfill
\begin{center}
{\bf\Large Quantum SU$(2|1)$ supersymmetric\\

\vspace{0.2cm}
 $\mathbb{C}^N$ Smorodinsky--Winternitz system}
\end{center}
\vspace{1.5cm}

\begin{center}
{\large\bf Evgeny Ivanov${\,}^{a),\,b)}$, Armen Nersessian${\,}^{c),\,a),\,d)}$, Stepan Sidorov${\,}^{a)}$}
\end{center}
\vspace{0.4cm}

\centerline{${\,}^{a)}$ \it Bogoliubov Laboratory of Theoretical Physics, JINR, 141980 Dubna, Moscow Region, Russia}
\vspace{0.2cm}
\centerline{${\,}^{b)}$ \it Moscow Institute of Physics and Technology, 141700 Dolgoprudny, Moscow Region, Russia}
\vspace{0.2cm}
{\centerline{${\,}^{c)}$ \it Yerevan Physics Institute, 2 Alikhanian Brothers St., 0036, Yerevan, Armenia}
\vspace{0.2cm}
\centerline{${\,}^{d)}$ \it Institute of Radiophysics and Electronics, Alikhanian 1, Ashtarak, 0203, Armenia}}
\vspace{0.3cm}

\centerline{eivanov@theor.jinr.ru, arnerses@yerphi.am, sidorovstepan88@gmail.com}
\vspace{0.2cm}

\vspace{2cm}

\par

{\abstract
\noindent We study quantum properties of SU$(2|1)$ supersymmetric (deformed ${\cal N}=4$, $d=1$ supersymmetric)
extension of the superintegrable Smorodinsky--Winternitz system on a complex Euclidian space $\mathbb{C}^N$.
The full set of wave functions is constructed and the energy spectrum is calculated. It is shown that SU$(2|1)$ supersymmetry
implies the bosonic and fermionic states to belong to separate energy levels, thus exhibiting the ``even-odd'' splitting of the spectra.
The superextended hidden symmetry operators are also defined and their action on SU$(2|1)$ multiplets of the wave functions is given. An equivalent
description of  the same system in terms of superconformal SU$(2|1,1)$ quantum mechanics is considered and a new representation of the hidden
symmetry generators in terms of the SU$(2|1,1)$ ones is found.}

\vfill{}


\end{titlepage}

\section{Introduction}
The models of supersymmetric mechanics were initially introduced as toy models for supersymmetric field theories \cite{W1}.
Respectively, their study was mainly limited to the models exhibiting one-dimensional Poincar\'e supersymmetry defined by the relations
\be
\{Q^A, Q^B\} = 2 \delta^{AB}H, \quad [H, Q^A] = 0, \qquad A = 1, \ldots {\cal N},\label{Nsusy}
\ee
where $Q^A$ are ${\cal N}$ real supercharges and $H$ is the Hamiltonian.

Some decade ago, there started a wide activity related to the study of field-theoretical models with the ``rigid supersymmetry on curved superspaces''
(see, e.g., \cite{FS}). These studies motivated  two of us (E.~I. and S.~S.) to investigate the one-dimensional ({\it i.e.} mechanical)
analogs of these theories \cite{DSQM}, \cite{SKO}.
The main subjects of our interest were systems in which $d=1$, $\mathcal{N}=4$ Poincar\'e supersymmetry is deformed by a mass-dimension parameter $m$ into $su(2|1)$ supersymmetry,
with the following non-vanishing (anti)commutators \footnote{These $su(2|1)$ superalgebra relations differ from those in \cite{ClSW}
by rescaling $Q\rightarrow \sqrt{2}\,Q$.
We mostly follow the conventions of \cite{SKO}.}:
\bea
    &&\left\lbrace Q^i,\bar{Q}_j \right\rbrace = 2m\,I^i_j +2\delta^i_j\,{\cal H},\qquad
    \left[I^i_j,  I^k_l\right] = \delta^k_j\,I^i_l - \delta^i_l\,I^k_j\,,\nn
    &&\left[I^i_j, \bar{Q}_{l}\right] = \frac{1}{2}\,\delta^i_j\,\bar{Q}_{l}-\delta^i_l\,\bar{Q}_{j}\,,\qquad \left[I^i_j, Q^{k}\right]
    = \delta^k_j\,Q^{i} - \frac{1}{2}\,\delta^i_j\,Q^{k},\nn
    &&\left[{\cal H}, \bar{Q}_{l}\right]=\frac{m}{2}\,\bar{Q}_{l}\,,\qquad \left[{\cal H}, Q^{k}\right]=-\,\frac{m}{2}\,Q^{k}.\label{su21}
\eea
Here, the  generator ${\cal H}$ is the Hamiltonian (coinciding with the ${\rm U}(1)$ generator), while the remaining bosonic generators
$I^i_j$ ($i=1,2$) define SU(2) R-symmetry.

As one of the results of this study, it was observed that the so-called ``weak $\mathcal{N}=4$ supersymmetric K\"ahler oscillator'' models (which are the particle models living
on K\"ahler spaces and interacting with the specific potential field and a constant magnetic field) suggested earlier in \cite{BelNer1,BelNer2} supply nice examples of SU$(2|1)$
supersymmetric mechanics \footnote{Another version of this kind of supersymmetric mechanics was studied in \cite{smilga}, with SU$(2|1)$ termed  as ``weak  supersymmetry''.}, and they
can be reproduced from the SU$(2|1)$ superfield approach worked out in \cite{SKO}. This observation further entailed the construction of a few novel SU$(2|1)$
supersymmetric superintegrable oscillator-like models specified by the interaction with a constant magnetic field. They include superextensions of isotropic oscillators
on $\mathbb{C}^N$ and $\mathbb{CP}^N$ \cite{BelNer1}, \cite{qcpn}, as well as of $\mathbb{C}^N$ Smorodinsky--Winternitz system in the presence of a constant
magnetic field \cite{Shmavonyan} and $\mathbb{CP}^N$ Rosochatius system \cite{Rosochatius}.

In a recent paper \cite{ClSW} we noticed that switching on an interaction with a constant magnetic field in the ordinary $\mathcal{N}=4$ supersymmetric mechanics on K\"ahler manifold
breaks either $\mathcal{N}=4$ supersymmetry or isometries of the initial bosonic system, including their hidden symmetries (cf. \cite{nonlinear}). We demonstrated
that this drawback can be overcome by performing, instead of the $d=1$ Poincar\'e supersymmetrization, its deformed variant, {\it i.e.} SU$(2|1)$ supersymmetrization.
Examining the SU$(2|1)$ supersymmetric models listed above, we observed that the SU$(2|1)$ supersymmetrization preserves all kinematical symmetries
of the initial systems, and, in a number of cases, also the hidden ones (the explicit expressions for the ``super-counterparts'' of the hidden symmetry generators
were constructed so far for $\mathbb{C}^N$ oscillator and $\mathbb{C}^N$ Smorodinsky--Winternitz system in the presence of a constant magnetic field).
It should be pointed out that the hidden symmetries play an important role in the quantum domain: e.g., in the standard supersymmetric quantum mechanics
they amount to an additional degeneracy of the (higher) energy levels as compared to $\mathcal{N}$-Poincar\'e supersymmetry which enhances it by the factor $2^{\mathcal{N}/2}$.

Keeping in mind these reasonings, it is desirable to better understand, on the concrete examples, how SU$(2|1)$ supersymmetry affects the energy spectrum
of the systems considered. The basic peculiar feature of the supersymmetric K\"ahler oscillator models is that their Hamiltonian is in fact identified
with the U(1)-generator in \eqref{su21}. As was shown in \cite{SKO}, an extra U$(1)$ R-charge (``fermionic number'') can be introduced in addition
to the Hamiltonian in this class of models  only in the limit $m=0$. One can expect that the impossibility to separate the Hamiltonian from the fermionic number
in such models at $m\neq 0$
could have an essential impact on the structure of the relevant energy spectra.

In the present paper we construct SU$(2|1)$ supersymmetric extension of the $\mathbb{C}^N$ Smorodinsky--Winternitz (in what follows, S.-W.)
quantum system in the presence of a constant magnetic field
as an instructive and simple example of the general K\"ahler oscillator models. In particular, we focus on the interplay of SU$(2|1)$ supersymmetry and hidden symmetry in forming the energy
spectrum of this system. An analogous analysis of its purely bosonic sector was performed in \cite{Shmavonyan}. An interesting unique feature of the
SU$(2|1)$ $\mathbb{C}^N$ S.-W. model as compared to other models of SU$(2|1)$ supersymmetric K\"ahler oscillators\footnote{This feature is shared by
SU$(2|1)$ $\mathbb{C}^N$ isotropic oscillator model, the Hamiltonian of which is a particular case of SU$(2|1)$ $\mathbb{C}^N$ S.-W. Hamiltonian corresponding
to the choice $g_a = 0$ in \eqref{SW} and \eqref{kahler} below.}
is its implicit superconformal SU$(2|1,1)$ symmetry. At the classical level this follows from the results of ref. \cite{ISTconf},
while here we extend the correspondence with a complex SU$(2|1,1)$ superconformal mechanics to the quantum domain.

As a preamble, let us remind that the $\mathbb{C}^N$ S.-W. system considered in  \cite{Shmavonyan} \footnote{Originally, S.-W. system was formulated
on the Euclidean $\mathbb{R}^N$ space \cite{SW1}.} amounts to a sum of $N$ two-dimensional isotropic oscillators
with a ring-shaped potential interacting with a constant magnetic field (orthogonal to the plane). It is defined by the Hamiltonian \cite{Shmavonyan}
\be
    {H} = \sum_{a=1}^N H_a\,,\quad H_a=
\bar{\pi}_a\pi_a+\frac{g^2_a}{4z^a\bar z^a}+\omega^2 z^a\bar{z}^a,\qquad \omega:=|\omega|\,,\label{SW}
\ee
with
\be
    \left[\pi_a, z^b\right]=-i\delta^b_a\,,\quad \left[\pi_a,\bar{\pi}_b\right]=  \delta_{a\bar b}\,B,
\ee
where $B$ is a constant strength of the magnetic field. Besides the standard Liouville integrals of motion, the model possesses the additional
ones  generated by the so-called Uhlenbeck tensor, and thus provides an example of superintegrable system. The $\mathbb{C}^N$ S.-W.
system interacting with a constant magnetic field belongs to the class of K\"ahler oscillators with the following K\"ahler potential \cite{ClSW}
\be
    K=\sum_{a=1}^N\left(z^a\bar{z}^{a}+\frac{i{g}_a}{\omega}\log{z^a}-\frac{i{ g}_a}{\omega}\log{{\bar z}^a}\right).\label{kahler}
\ee
It admits SU$(2|1)$ supersymmetric extension given by the superalgebra \eqref{su21}, with the deformation parameter
\be
    m=\sqrt{4\omega^2 +B^2}\,.\label{m}
\ee
This extension, at the classical level, was described in \cite{ClSW}. It is a particular case of general SU$(2|1)$ mechanics models based on
the multiplets ${\bf (2, 4, 2)}$ \cite{SKO}.\\

It is convenient to summarize here at once the basic results of the present paper.

\begin{itemize}

\item The energy spectrum of the deformed SU$(2|1)$ supersymmetric system constructed reveals an interesting feature: SU$(2|1)$ supersymmetry
gives rise to the separation of bosonic and fermionic states in the spectrum. Bosonic states are associated with the even levels, while all fermionic states with the odd ones.
So the energy spectrum exhibits ``even-odd'' feature: the adjacent bosonic and fermionic states carry different energies shifted by half-integer numbers.
The intrinsic reason for such a splitting is that the Hamiltonian in \eqref{su21} hides in itself
the fermionic number operator $F$ and so does not commute with the supercharges. The ground state belongs to a non-singlet representation of SU$(2|1)$, so the
latter is spontaneously broken. These features are in contrast with the standard supersymmetric mechanics models where there is a degeneracy between the
fermionic and bosonic wave functions.

\item The system exhibits superconformal SU$(2|1,1)$ symmetry with the central charge given by the sum of the generators of kinematical
U(1) symmetries. The SU$(2|1)$ supersymmetric Hamiltonian
$\mathcal{H}$ can be split into a sum of superconformal Hamiltonian
and a central charge generator accounting for the constant external
magnetic field. The whole set of wave functions is closed under the
action of the superconformal algebra $su(2|1,1)$, which can be
treated as a spectrum-generating algebra of the model considered.

\item Due to an additional hidden symmetry generated by the proper superanalog of Uhlenbeck tensor commuting with the Hamiltonian, the spectrum of quantum SU$(2|1)$
$\mathbb{C}^N$ S.-W. system reveals an extra degeneracy. The superextended Uhlenbeck tensor is shown to admit a nice representation
in terms of bilinears of the  SU$(2|1,1)$ generators.

\item Furthermore, a generalization of the  super Uhlenbeck tensor was found, such that it commutes with the full set of the generators of the properly rotated
SU$(2|1)$ supergroup. It is also expressed through  SU$(2|1,1)$ generators and gives rise to an additional degeneracy of the eigenvalues
of the SU$(2|1)$ Casimirs.

\end{itemize}

The paper is organized as follows. In Section 2 we formulate SU$(2|1)$ supersymmetric extension of
the superintegrable $\mathbb{C}^1$ S.-W. quantum system and determine its energy spectrum and wave functions.
We also analyze the space of quantum states of this supersymmetric system from the standpoint of the SU$(2|1)$
representation theory. In Section 3 we show that the system considered admits an equivalent description in terms of some quantum superconformal SU$(2|1,1)$ mechanics.
In Section 4 we define the SU$(2|1)$ supersymmetric $\mathbb{C}^N$ S.-W. quantum system as a sum of
$N$ copies of the $\mathbb{C}^1$-systems. We reveal its hidden symmetry given by the supersymmetric counterpart of the Uhlenbeck tensor and show that it is responsible
for an additional degeneracy of the spectrum of the SU$(2|1)$ Casimir operators. A new representation for the superextended Uhlenbeck tensor in terms of
the superconformal SU$(2|1,1)$ generators is found. The summary and outlook are the contents of Section 5.
In Appendices A and B we present some details related to the non-linear algebra of hidden symmetries.
In Appendix C, it is briefly discussed how conformal SL(2,R) symmetry manifests itself in the spectrum of quantum bosonic $\mathbb{C}^N$ S.-W. system.

\section{Supersymmetric $\mathbb{C}^1$ Smorodinsky--Winternitz system}
We proceed from the SU$(2|1)$ supersymmetric S.-W. system on the complex Euclidian space $\mathbb{C}^1$.
The quantum Hamiltonian constructed according to the generic prescription of \cite{SKO,ClSW} for the specific $N=1$ K\"ahler potential \eqref{kahler},
is defined by the expression,
\be
    \mathcal{H}=\bar{\pi}\pi+\frac{g^2}{4z\bar z} + \omega^2 z\bar{z}\,
 +ig\left(\frac{\xi_k\xi^k}{4z^2}-\frac{\bar{\xi}^k\bar{\xi}_k}{4\bar{z}^2}\right)+\frac{B}{2}\,\xi^k \bar{\xi}_k\,, \label{qH}
\ee
with the (anti)commutators\footnote{For reader's convenience we quote here the dimensions of various quantities (in the length units):
$$
[z] =cm^{1/2}\,, \; [\xi_k] = [g] = cm^0\,, \; [B] = [m] = [\omega] = [\mathcal{H}] = cm^{-1}\,.
$$}
\be
    \left[\pi,z\right]=-i,\qquad\left[\bar{\pi},\bar{z}\right]=-i,\qquad\left[\pi,\bar{\pi}\right]=B,\qquad
    \left\lbrace\xi^{i} ,\bar{\xi}_{j}\right\rbrace = \xi^{i}\bar{\xi}_{j}+\bar{\xi}_{j}\xi^{i} = \delta^i_j\,.
\ee
The operators of bosonic momenta are represented in the standard form
\be
    \pi=-\,i\left(\partial_{z} + \frac{B}{2}\,\bar{z}\right),\quad
    \bar{\pi}=-\,i\left(\partial_{\bar{z}} - \frac{B}{2}\,z\right),
\ee
while the fermionic operators $\xi^i$, $\bar\xi_i$ can be represented as a $4\times 4$ Euclidian gamma-matrices
acting on four-component wave functions. However, we prefer a more formal, though equivalent representation, see, e.g., \cite{smilga85}.
Namely, we consider one-component wave functions depending on $z$, $\bar z$, $\xi^i$ (with $\xi^i, i=1,3\,,$ being a doublet of  Grassmann variables),
and represent $\bar{\xi}_j$ as a differential operator,
\be
    \bar{\xi}_j=\partial/\partial \xi^j.
\ee
Respectively, the wave functions contain two bosonic components,  $\psi\left(z,\bar z\right)$,
$\xi_k\xi^k\psi'\left(z,\bar z\right)$, and two fermionic ones,
$\Psi^i=\xi^i\psi''\left(z,\bar z\right)$.

In what follows, it will be convenient to equivalently replace the parameters $B$ and $\omega$ by $\lambda$ and $m$, where $m$ is a mass-dimension contraction parameter defined in \eqref{m},
and $\lambda$ is a dimensionless angle-type parameter defined by the relations
 \be
    B = m\cos{2\lambda}\,,\qquad \omega = \frac{m\sin{2\lambda}}{2}\,,\qquad m = \sqrt{4\omega^2 + B^2} >0,\quad \lambda\in\left[0,\pi/2\right).
\label{Bomega}
\ee
In the new notation, the  Hamiltonian \eqref{qH} is rewritten as
\bea
    {\cal H}=-\,\partial_{\bar{z}}\partial_{z}+\frac{m\cos{2\lambda}}{2}\,L + \frac{m^2z\bar{z}}{4}+\frac{g^2}{4z\bar{z}}+ig\left(\frac{\xi_k\xi^k}{4z^2}
    -\frac{\bar{\xi}^k\bar{\xi}_k}{4\bar{z}^2}\right),\label{qH2}
\eea
where $L$ is the angular momentum (or U(1)) operator
\bea
    L = z\partial_{z}-\bar{z}\partial_{\bar{z}}+\xi^k\bar{\xi}_k-1\,. \label{angular}
\eea
The Hamiltonian \eqref{qH2} is manifestly invariant under U(1)-transformation
$z\to {\rm e}^{i\kappa}z$, $\xi^i\to {\rm e}^{i\kappa}\xi^i$ generated by this operator.
The rest of SU$(2|1)$ generators is expressed as
\bea
    &&Q^i = -\sqrt{2}\,i\left[\cos{\lambda}\;\xi^{i}\left(\partial_{z} + \frac{m}{2}\,\bar{z}\right)
      -\sin{\lambda}\;\bar{\xi}^i\left(\partial_{\bar{z}} + \frac{m}{2}\,z\right)\right]+\frac{g}{\sqrt{2}}\left(\frac{\sin{\lambda}\;\xi^{i}}{z}
      +\frac{\cos{\lambda}\;\bar{\xi}^i}{\bar{z}}\right),\nn
       &&\bar{Q}_{j} = -\sqrt{2}\,i\left[\cos{\lambda}\;\bar{\xi}_{j}\left(\partial_{\bar{z}} - \frac{m}{2}\,z\right) +\sin{\lambda}\;\xi_j\left(\partial_{z} - \frac{m}{2}\,\bar{z}\right)\right]+\frac{g}{\sqrt{2}}\left(\frac{\sin{\lambda}\;\bar{\xi}_j}{\bar{z}}-\frac{\cos{\lambda}\;\xi_{j}}{z}\right),\nn
    &&I^i_j = \xi^{i}\bar{\xi}_{j}  -\frac{{\delta_j^i}}{2}\,\xi^{k} \bar{\xi}_{k}\,.\label{I}
\eea
It is straightforward to check that they satisfy the $su(2|1)$ superalgebra relations \eqref{su21}.
The operator $L$ commutes with all $su(2|1)$ generators, $\left[L,\mathcal{H}\right]=\left[L,Q^i\right]=\left[L,I^i_j\right]=0$,
{\it i.e.} it can be interpreted as a {\it central charge}. The SU(2) generators $I^i_j$ commute with the Hamiltonian \eqref{qH2},
implying that this SU(2) is realized on the quantum states as an exact symmetry.

\subsection{Wave functions and spectrum}
We define super wave functions as a $\xi^i$-expansion of the wave functions depending on $(z, \bar z, \xi^i)$,
with $\bar{\xi}_j$ being an annihilation operator. This expansion amounts to the fermionic wave function $\Psi^i \sim \xi^i \psi''\left(z,\bar{z}\right)$ and two bosonic
wave functions $\psi\left(z,\bar z\right)$,
$\xi_k\xi^k\psi'\left(z,\bar z\right)$. As distinct from the fermionic function, the bosonic ones  are not eigenstates of the Hamiltonian \eqref{qH2}.
The correct eigenstates are represented as their proper combinations:
\bea
    \Omega \sim \psi\left(z,\bar z\right)+\xi_k\xi^k\psi'\left(z,\bar z\right)\label{GenStruc}
\eea
(see eq. \eqref{omega12} below).

The simplest way to solve the eigenvalue problem is to firstly consider the action of \eqref{qH2}
on the fermionic wave functions $\Psi^i \sim \xi^i \psi''\left(z,\bar{z}\right)$. On the bosonic factor $\psi''\left(z,\bar{z}\right)$
the Hamiltonian \eqref{qH2} acts exactly as the bosonic one [cf. \eqref{SW}], and this action reads
\be
    {\cal H}_{0} = -\,\partial_{\bar{z}}\partial_{z}+\frac{m\cos{2\lambda}}{2}\,\left(z\partial_{z}-\bar{z}\partial_{\bar{z}}\right) + \frac{m^2z\bar{z}}{4}+\frac{g^2}{4z\bar{z}}\,.
\label{C1SW}
\ee
After solving the eigenvalue problem as in the pure bosonic case \cite{Shmavonyan}, the fermionic states are written as
\bea
    \Psi^i_{(n , l)} =\xi^i\left(n-1\right)!\,\left(z\bar{z}\right)^{\frac{\tilde{l}}{2}}\left(z/\bar{z}\right)^{\frac{l}{2}} e^{-\frac{mz\bar{z}}{2}}L^{(\tilde{l}\,)}_{n-1}\left(mz\bar{z}\right),\qquad
    \tilde{l} = \sqrt{l^2 + g^2}\,,\label{fermionicwf}
\eea
where $L^{(\tilde{l}\,)}_{n-1}$ are generalized Laguerre polynomials, {\it i.e.} $n=1, 2, 3 \ldots$ is a positive integer
\footnote{One could work with the Laguerre polynomials $L^{(\tilde{l}\,)}_{n^\prime}\left(mz\bar{z}\right)$ where $n^\prime=0, 1, 2 \ldots$ ,
but for further convenience we deal with $n=n^\prime + 1$, such that $n=1, 2, 3 \ldots$ (see the next Subsection).}. The integer number $l$
is an eigenvalue of the angular-momentum operator $L$,
\bea
    L\,\Psi^i_{(n,l)} = l\,\Psi^i_{(n,l)}\,,\qquad l=0, \pm 1, \pm2  \ldots .
\eea
The energy spectrum of the fermionic states is directly calculated to be
\bea
    {\cal H}_{0}\,\Psi^i_{(n,l)} = E_{(n,l)}\Psi^i_{(n,l)}\,,\qquad E_{(n , l)}=m\left[n+\frac{1}{2}\left(\tilde{l}+l\cos{2\lambda}-1\right)\right].\label{Flevels}
\eea
The double degeneracy of the fermionic levels is due to the unbroken $su(2)\subset su(2|1)$ with respect
to which the wave function \eqref{fermionicwf} transforms as
a doublet.

The bosonic wave functions can be now obtained by action of the supercharges on the fermionic wave function $\Psi^i$\,:
\bea
    Q^i\Psi^j_{(n,l)} = \varepsilon^{ij}\sqrt{2m}\,\Omega^{-}_{(n,l)}\,,\qquad \bar{Q}_{j}\Psi^i_{(n,l)} = \delta^i_j\sqrt{2m}\,\Omega^{+}_{(n,l)}\,.\nn
\eea
They are explicitly expressed as
\bea
    \Omega^{-}_{(n,l)}&=&\frac{\left(n-1\right)!}{2\sqrt{m}}\left(z\bar{z}\right)^{\frac{\tilde{l}}{2}}\left(z/\bar{z}\right)^{\frac{l}{2}} e^{-\frac{mz\bar{z}}{2}}\bigg[\frac{1}{\bar{z}}\left(g\cos{\lambda}-il\sin{\lambda}+i\tilde{l}\sin{\lambda}\right)L^{(\tilde{l}\,)}_{n-1}\left(mz\bar{z}\right)\nn
    &&+\,\frac{1}{2z}\,\xi_k\xi^k\left(g\sin{\lambda}-il\cos{\lambda}-i\tilde{l}\cos{\lambda}\right)L^{(\tilde{l}\,)}_{n-1}\left(mz\bar{z}\right)\nn
    &&+\,im\left(2z\sin{\lambda} - \bar{z}\,\xi_k\xi^k\cos{\lambda}\right)\left[L^{(\tilde{l}\,)}_{n-1}\left(mz\bar{z}\right)-L^{(\tilde{l}+1)}_{n-1}\left(mz\bar{z}\right)\right]\bigg],\nn
    \Omega^{+}_{(n,l)}&=&\frac{\left(n-1\right)!}{2\sqrt{m}}\left(z\bar{z}\right)^{\frac{\tilde{l}}{2}}\left(z/\bar{z}\right)^{\frac{l}{2}} e^{-\frac{mz\bar{z}}{2}}
    \bigg[\frac{1}{\bar{z}}\left(g\sin{\lambda}+il\cos{\lambda}-i\tilde{l}\cos{\lambda}\right)L^{(\tilde{l}\,)}_{n-1}\left(mz\bar{z}\right)\nn
    &&-\,\frac{1}{2z}\,\xi_k\xi^k\left(g\cos{\lambda}+il\sin{\lambda}+i\tilde{l}\sin{\lambda}\right)L^{(\tilde{l}\,)}_{n-1}\left(mz\bar{z}\right)\nn
    &&+\,im\left(2z\cos{\lambda} + \bar{z}\,\xi_k\xi^k\sin{\lambda}\right)L^{(\tilde{l}+1)}_{n-1}\left(mz\bar{z}\right)\bigg], \qquad L\,\Omega^{\pm}_{(n,l)}
    = l\,\Omega^{\pm}_{(n,l)}\,. \label{omega+-}
\eea
On the other hand, the action of supercharges on the bosonic wave functions yields the original fermionic wave functions
$\Psi^i$:
\bea
    &&Q^i\,\Omega^{+}_{(n,l)} = \left[n + \frac{1}{2}\left(\tilde{l}
 + l\cos{2\lambda}\right)\right]\sqrt{2m}\,\Psi^{i}_{(n,l)}\,,\qquad \bar{Q}_{j}\,\Omega^{+}_{(n,l)}=0,\nn
    &&\bar{Q}^{j}\,\Omega^{-}_{(n,l)} = -\left[n - 1 + \frac{1}{2}\left(\tilde{l}
    + l\cos{2\lambda}\right)\right]\sqrt{2m}\,\Psi^j_{(n,l)}\,,\qquad Q^i\,\Omega^{-}_{(n,l)} =0\,.
\eea
Using these expressions, we derive the energy spectrum for the bosonic states $\Omega^{\pm}_{(n,l)}$
and find that it differs from that for the fermionic states,
\be
    {\cal H}\,\Omega^{\pm}_{(n,l)}=\left(E_{(n,l)}\pm\frac{m}{2}\right)\Omega^{\pm}_{(n,l)}\,.\label{Blevels}
\ee
Thus SU$(2|1)$ supersymmetry creates additional energy levels, with the bosonic and fermionic states being separated.

Note that the discrete energy spectrum bounded from below in the model under consideration
is ensured by the oscillator term $\sim m^2$ in \eqref{C1SW}. So the limit $m=0$
cannot be taken in the quantum case (see also Section 3).

\subsection{SU$(2|1)$ representations}
The SU$(2|1)$ representations are specified by the eigenvalues of Casimir operators \cite{repres}
\bea
    \label{C2}
    &C_2& = \frac{{\cal H}^2}{m^2}-\frac{I^i_jI^j_i}{2}+\frac{Q^k\bar{Q}_k-\bar{Q}_kQ^k}{4m}\,,\\
    &C_3& = \left(C_2+\frac{1}{2}\right)\frac{{\cal H}}{m}+\frac{1}{8m^2}\left(\delta^i_j{\cal H}-m\,I^i_j\right)\left(Q^j\bar{Q}_i-\bar{Q}_iQ^j\right).\label{C3}
\eea
On the quantum states $\Psi^j_{(n,l)}\,, \Omega^{\pm}_{(n,l)}$ these eigenvalues are:
\bea
    C_2 &=& \left[n + \frac{1}{2}\left(\tilde{l} + l\cos{2\lambda}\right)\right]\left[n + \frac{1}{2}\left(\tilde{l}  + l\cos{2\lambda}\right)-1\right],\nn
    C_3 &=& \left[n + \frac{1}{2}\left(\tilde{l}  + l\cos{2\lambda}-1\right)\right]C_2\,.\label{eigenvalues}
\eea
The quantum number $n$ uniquely defines SU$(2|1)$ representation for a fixed $l$.
Casimir operators  take non-zero eigenvalues \eqref{eigenvalues} on all states. It means that all SU$(2|1)$ representations are typical with the simplest four-fold
degeneracy \footnote{This degeneracy is with respect to
the Casimir operators, not to the Hamiltonian.}. Respectively, the SU$(2|1)$ multiplets are encompassed by the sets
$$
\left\lbrace\Psi^j_{(n,l)}\,, \,\Omega^{-}_{(n,l)}\,, \,\Omega^{+}_{(n,l)}\right\rbrace
$$
and constitute infinite-dimensional ``towers'' characterized by all values of the angular-momentum $L$, $l=0, \pm 1, \pm2  \ldots $ for a fixed $n$.

The minimal energy for a fixed $l$ is always positive and it corresponds to the bosonic state $\Omega^{-}_{(1,l)}$\,:
\bea
    \tilde{l}=\sqrt{l^2 + g^2}>l\quad \Rightarrow \quad E_{\rm min}=\frac{m}{2}\left(\tilde{l} + l\cos{2\lambda}\right)>0.
\eea
Therefore, the ground states for each $l$ belongs to a non-trivial four-fold SU$(2|1)$ multiplet. This means  that SU$(2|1)$ supersymmetry is spontaneously broken
at any $l$.

\begin{figure}[t]
\begin{center}
\begin{picture}(270,210)

\put(80,5){\line(0,1){190}}
\put(80,185){\vector(0,1){10}}

\put(90,30){\line(1,0){180}}
\put(90,90){\line(1,0){180}}
\put(90,150){\line(1,0){180}}
\put(90,60){\line(1,0){180}}
\put(90,120){\line(1,0){180}}
\put(90,180){\line(1,0){180}}

\put(250,30){\circle*{11}}
\put(110,90){\circle*{11}}
\put(110,150){\circle*{11}}

\multiput(160,55)(30,0){2}{{\LARGE $\times$}}
\multiput(160,115)(30,0){2}{{\LARGE $\times$}}
\multiput(160,175)(30,0){2}{{\LARGE $\times$}}

\put(250,90){\circle*{11}}
\put(250,150){\circle*{11}}

\put(15,26){$E_{(1,l)} - m/2$}
\put(30,56){$E_{(1,l)}$}
\put(15,86){$E_{(1,l)} + m/2$}
\put(30,116){$E_{(2,l)}$}
\put(15,146){$E_{(2,l)} + m/2$}
\put(30,176){$E_{(3,l)}$}
\put(60,195){${\cal H}$}
\put(175,0){$\Psi^i_{(n,l)}$}
\put(105,0){$\Omega^+_{(n,l)}$}
\put(245,0){$\Omega^-_{(n,l)}$}

\put(130,80){\line(3,-1){40}}
\put(130,80){\vector(-3,1){10}}
\put(166,68){\vector(3,-1){10}}
\put(120,75){\tiny $\bar{Q}_{j}$}
\put(170,70){\tiny $Q^{i}$}

\put(130,140){\line(3,-1){40}}
\put(130,140){\vector(-3,1){10}}
\put(166,128){\vector(3,-1){10}}
\put(120,135){\tiny $\bar{Q}_{j}$}
\put(170,130){\tiny $Q^{i}$}

\put(130,200){\line(3,-1){40}}
\put(130,200){\vector(-3,1){10}}
\put(166,188){\vector(3,-1){10}}
\put(120,195){\tiny $\bar{Q}_{j}$}
\put(170,190){\tiny $Q^{i}$}

\put(185,113){\line(3,-1){45}}
\put(230,98){\vector(3,-1){10}}
\put(194,110){\vector(-3,1){10}}
\put(230,100){\tiny $Q^{i}$}
\put(180,165){\tiny $\bar{Q}_{j}$}

\put(185,173){\line(3,-1){45}}
\put(230,158){\vector(3,-1){10}}
\put(194,50){\vector(-3,1){10}}
\put(230,160){\tiny $Q^{i}$}
\put(180,105){\tiny $\bar{Q}_{j}$}

\put(185,53){\line(3,-1){45}}
\put(230,38){\vector(3,-1){10}}
\put(194,170){\vector(-3,1){10}}
\put(230,40){\tiny $Q^{i}$}
\put(180,45){\tiny $\bar{Q}_{j}$}

\end{picture}
\end{center}
\caption{Degeneracy of energy levels with a fixed number $l$. Bosonic (circles) and fermionic (crosses)
states of the 4-fold SU$(2|1)$ multiplet, characterized by the number $n$, fill the energy levels $E_{(n,l)} \pm m/2$ and $E_{(n,l)}$.}
\label{figure1}
\end{figure}
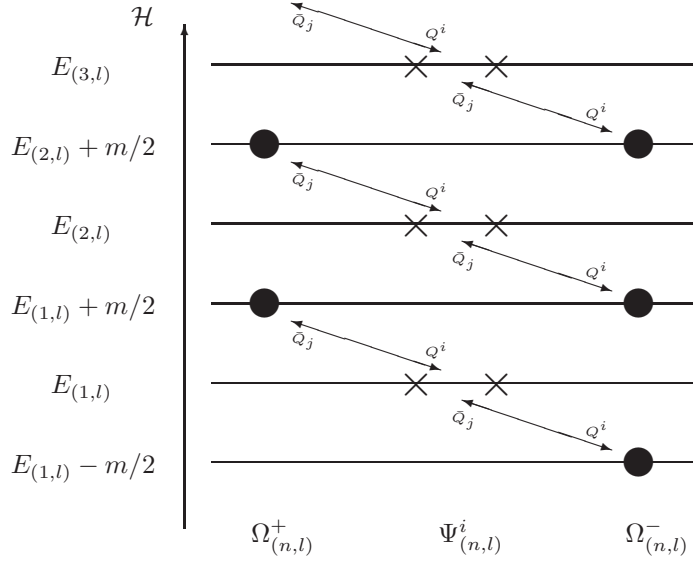

The quantum states within the given SU$(2|1)$ multiplet occupy different energy levels according to \eqref{Flevels} and \eqref{Blevels}.
Since the supercharges do not commute with the Hamiltonian \eqref{qH2}, they are capable to decrease and increase its eigenvalues.
One can designate levels occupied by bosonic states
as even levels (including the zeroth level for the ground state), then the  fermionic states occupy odd levels. As distinct from the standard $d=1$ Poincar\'e supersymmetry, there is
no any degeneracy between these two types of the levels. The relevant picture is drawn on Figure \ref{figure1}.

One can see that the bosonic states $\Omega^{+}_{(n,l)}$ and $\Omega^{-}_{(n+1,l)}$ have the same energy though they belong to different SU$(2|1)$ representations,
\bea
    E_{(n,l)}+\frac{m}{2} = E_{(n + 1,l)} - \frac{m}{2}=m\left[n+\frac{1}{2}\left(\tilde{l}+l\cos{2\lambda}\right)\right],\qquad n=1, 2, 3 \ldots .\label{+-}
\eea
For example, the energy level $E_{(1,l)}+m/2$ on Figure \ref{figure1} can be alternatively denoted as $E_{(2,l)} - m/2$.
The hidden symmetry responsible for this degeneracy will be presented in the next section. As was already mentioned, the two-fold degeneracy of fermionic states
is due to the ${\rm SU}(2)$ generators \eqref{I} acting on the fermionic variables $\xi^i$.

The wave functions $\Omega^{+}_{(n,l)}$ and $\Omega^{-}_{(n+1,l)}$ for $n=1, 2, 3 \ldots $ can be represented as
\bea
    &&\Omega^{-}_{(n+1,l)}=\frac{\left(\tilde{l}+n\right)}{g\,\tilde{l}}\left[\left(\tilde{l}+l\right)\cos{\lambda}+ig\sin{\lambda}\right]\Omega^{1}_{(n,l)}
    +\frac{n}{2\,g\,\tilde{l}}\left[\left(\tilde{l}-l\right)\cos{\lambda}-i\sin{\lambda}\right]\Omega^{2}_{(n,l)}\,,\nn
    &&\Omega^{+}_{(n,l)}=\frac{\left(\tilde{l}+n\right)}{g\,\tilde{l}\,n}\left[\left(\tilde{l}+l\right)\sin{\lambda}-ig\cos{\lambda}\right]\Omega^{1}_{(n,l)}
    +\frac{1}{2\,g\,\tilde{l}}\left[\left(\tilde{l}-l\right)\sin{\lambda}+i\cos{\lambda}\right]\Omega^{2}_{(n,l)}\,,\label{omega-omega12}
\eea
where
\bea
    &&\Omega^{1}_{(n,l)}= \frac{n!}{2\sqrt{m}}\left(z\bar{z}\right)^{\frac{\tilde{l}}{2}}\left(z/\bar{z}\right)^{\frac{l}{2}} e^{-\frac{mz\bar{z}}{2}}\left[\frac{1}{\bar{z}}\left(\tilde{l}-l\right)-\frac{ig}{2z}\,\xi_k\xi^k\right]L^{(\tilde{l}-1)}_{n}\left(mz\bar{z}\right),\nn
    &&\Omega^{2}_{(n,l)}= \frac{\left(n-1\right)!\,\sqrt{m}}{2}\left(z\bar{z}\right)^{\frac{\tilde{l}}{2}}\left(z/\bar{z}\right)^{\frac{l}{2}} e^{-\frac{mz\bar{z}}{2}}\left[2\left(\tilde{l}+l\right)z +ig\bar{z}\,\xi_k\xi^k\right]L^{(\tilde{l}+1)}_{n-1}\left(mz\bar{z}\right).
    \label{omega12}
\eea
These functions have just the structure \eqref{GenStruc}  and are eigenfunctions of the Hamiltonian \eqref{qH2} with the eigenvalues \eqref{+-}, but they are not
eigenfunctions of Casimirs \eqref{C2} and \eqref{C3}. The only exception is the wave function $\Omega^{1}_{(n,l)}$ which can be naturally continued to $n=0$ as
\bea
    \Omega^{1}_{(0,l)} &=& \frac{1}{2\sqrt{m}}\left(z\bar{z}\right)^{\frac{\tilde{l}}{2}}\left(z/\bar{z}\right)^{\frac{l}{2}} e^{-\frac{mz\bar{z}}{2}}\left[\frac{1}{\bar{z}}\left(\tilde{l}-l\right)-\frac{ig}{2z}\,\xi_k\xi^k\right].\label{omega0}
\eea
It is directly related to the lowest bosonic state $\Omega^{-}_{(1,l)}$\,\footnote{Note that the apparent singularities in $g$ and $l$ in eqs. \eqref{omega-omega12} and \eqref{01l}
are fake: no such singularities are present in the original expressions \eqref{omega+-} for $\Omega^{\pm}_{(n,l)}$.}:
\bea
    \Omega^{-}_{(1,l)}=\left[\frac{\left(\tilde{l}+l\right)\cos{\lambda}}{g}+i\sin{\lambda}\right]\Omega^{1}_{(0,l)}\,.\label{01l}
\eea
In the next section we will show that the state $\Omega^{1}_{(0,l)}$ can be interpreted as a singlet ground state with respect to some new
SU$(2|1)$ supercharges $\tilde{Q}$, $\bar{\tilde{Q}}\,$.

Let us summarize the basic peculiarities of the energy spectrum.\\

The system has two fermionic and two bosonic sets of iso-spectral states, with the energy levels given by \eqref{Flevels}
and \eqref{Blevels}, respectively, so that each energy level is doubly degenerated (excepting for the lowest level).
In terms of the ``physical'' parameters $B$, $\omega$ the energy spectrum reads
\bea
    E^{\sigma}_{(n , l)}=\sqrt{4\omega^2 +B^2}\left(n-\sigma+\frac{\sqrt{l^2+g^2}}{2}\right)+ \frac{Bl}{2}\,, \label{Flevels1}
\eea
with $n=1, 2, 3 \ldots ,\; l=0, \pm 1, \pm 2 \ldots ,\; \sigma=0, 1/2, 1,$
where $\sigma=0$ and $\sigma=1$ correspond to the bosonic states $\Omega^{+}_{(n,l)}$ and $\Omega^{-}_{(n,l)}$\,, respectively,
and $\sigma=1/2$ to the doublet of fermionic states.

\section{Superconformal symmetry}
In this section, following \cite{ISTconf}, we relate the generic supersymmetric $\mathbb{C}^1$ S.-W. system to the superconformal mechanics with the Hamiltonian
\bea
  \mathcal{H}_{\rm conf} = -\,\partial_{\bar{z}}\partial_{z}+\frac{m^2z\bar{z}}{4}+\frac{g^2}{4z\bar{z}}
   +ig\left(\frac{\xi_k\xi^k}{4z^2}-\frac{\bar{\xi}^k\bar{\xi}_k}{4\bar{z}^2}\right).
    \label{confH}
\eea
We will show that the Hamiltonians of these two systems differ by a central charge generator, and, as a consequence,
the Hamiltonian of conformal model inherits all the symmetries of the original Hamiltonian (and vice-versa).

As was proved in \cite{ISTconf}, the superconformal Hamiltonian of deformed supersymmetric mechanics should be an even function
of the deformation mass parameter $m$\,: $m \rightarrow -\,m$, $\mathcal{H}_{\rm conf} \rightarrow \mathcal{H}_{\rm conf}$\,.
In accord with this proposition we define the superconformal Hamiltonian \eqref{confH} as
\bea
\mathcal{H}_{\rm conf} = {\cal H} + \frac{m}{2}\,Z_1\,,\qquad{\rm with}\quad  Z_1 = -\cos{2\lambda}\;L.\label{redef}
\eea
Such a change of the Hamiltonian amounts to the effective elimination of the magnetic field
\footnote{With $B=0$, {\it i.e.} $\lambda=\pi/4$, the deformation parameter coincides with the frequency, and the central charge $Z_1$ vanishes:
\be
\lambda=\frac{\pi}{4}\quad\Rightarrow \quad B=0,\quad m=2\omega,\quad Z_1=0.\nonumber
\ee
Therefore, at this special choice of parameters the $\mathbb{C}^1$ S.-W. Hamiltonian \eqref{C1SW} just coincides with the superconformal Hamiltonian
\eqref{confH}, $\mathcal{H}=\mathcal{H}_{\rm conf}$\,.}.

One can equally choose the basis in which the conformal Hamiltonian is not deformed by the oscillator term,
\bea
    H_{\rm conf} = \mathcal{H}_{\rm conf} - \frac{m^2z\bar{z}}{4} = -\,\partial_{\bar{z}}\partial_{z}+\frac{g^2}{4z\bar{z}}
   +ig\left(\frac{\xi_k\xi^k}{4z^2}-\frac{\bar{\xi}^k\bar{\xi}_k}{4\bar{z}^2}\right).\label{confH0}
\eea
Then we introduce the dilatation generator $D$ and the generator of conformal boosts $K$:
\be
D:=\frac{i}{2}\left(z\partial_z+\bar z\partial_{\bar z}+1\right),\quad K:= z\bar{z}. \label{DefDK}
\ee
These generators, together with the Hamiltonian \eqref{confH0}, close on the conformal algebra $so(2,1) \sim sl(2,R)$ \cite{DFF}:
\bea
    [H_{\rm conf}\,,D]=iH_{\rm conf}\,,\quad [K, D]=-\,iK,\quad[H_{\rm conf}\,,K]=2iD.\label{so210}
\eea
The trigonometric type of (super)conformal mechanics involving the parameter $m$ is defined by the following linear combinations  \cite{ISTconf}:
\bea
\mathcal{H}_{\rm conf} = H_{\rm conf} + \frac{m^2}{4}\,K,\quad
T=H_{\rm conf}- \frac{m^2}{4}\,K - imD,\quad {\bar T}=H_{\rm conf}- \frac{m^2}{4}\,K + imD.\label{HKD}
\eea
The algebra \eqref{so210} is then rewritten as
\bea
    \left[\bar{T},T\right]=2m\,\mathcal{H}_{\rm conf}\,,\quad\left[\mathcal{H}_{\rm conf}\,, \bar{T}\right]=-\,m\,\bar{T},\quad \left[\mathcal{H}_{\rm conf}\,, T\right]=m\,T.
\label{so21}
\eea
One could come back to the original Hamiltonian $\mathcal{H}$ according to \eqref{redef}, but in this case the conformal algebra will be deformed by a central charge.
So, irrespective of whether we deal with superconformal symmetry or its bosonic limit, it is appropriate to use the Hamiltonian $\mathcal{H}_{\rm conf}$ containing
no magnetic field. Note that the first relation in \eqref{HKD} implies that  just $\mathcal{H}_{\rm conf}$ with $m^2\neq 0$, as opposed to $H_{\rm conf}$\,,
is the correct quantum Hamiltonian with the spectrum bounded from below, in accordance with the assertion in the pioneering paper \cite{DFF}.

The conformal algebra \eqref{so21} can be extended to the superconformal algebra in the following way. Applying the discrete transformation
$m\to -\,m$ to the supercharges $Q^i$ defined by \eqref{I} we obtain the new fermionic generators which can be identified
with the conformal supercharges, $S^i:=Q^i(-m)$:
\bea
    &&S^i = -\sqrt{2}\,i\left[\cos{\lambda}\;\xi^{i}\left(\partial_{z} - \frac{m}{2}\,\bar{z}\right)
      -\sin{\lambda}\;\bar{\xi}^i\left(\partial_{\bar{z}} - \frac{m}{2}\,z\right)\right]+\frac{g}{\sqrt{2}}\left(\frac{\sin{\lambda}\;\xi^{i}}{z}
      +\frac{\cos{\lambda}\;\bar{\xi}^i}{\bar{z}}\right),\nn
       &&\bar{S}_{j} = -\sqrt{2}\,i\left[\cos{\lambda}\;\bar{\xi}_{j}\left(\partial_{\bar{z}} + \frac{m}{2}\,z\right) +\sin{\lambda}\;\xi_j\left(\partial_{z} + \frac{m}{2}\,\bar{z}\right)\right]+\frac{g}{\sqrt{2}}\left(\frac{\sin{\lambda}\;\bar{\xi}_j}{\bar{z}}-\frac{\cos{\lambda}\;\xi_{j}}{z}\right).\nn
\eea
These new generators extend the superalgebra $su(2|1)$ to the centrally extended superalgebra $su(2|1,1)$\,:
\bea
    &&\left\lbrace Q^i,\bar{Q}_j \right\rbrace = 2m\,I^i_j - m\,\delta^i_j\,Z_1 + 2\delta^i_j\,\mathcal{H}_{\rm conf}\,,\qquad
    \left\lbrace S^i,\bar{S}_j \right\rbrace = -\, 2m\,I^i_j + m\,\delta^i_j\,Z_1 + 2\delta^i_j\,\mathcal{H}_{\rm conf}\,,\nn
    &&\left\lbrace S^i,\bar{Q}_j \right\rbrace = 2\delta^i_j\,T,\quad \left\lbrace Q^i,\bar{S}_j \right\rbrace = 2\delta^i_j\,\bar{T},\quad
    \left\lbrace Q^i,S^j\right\rbrace = m\,\varepsilon^{ij} Z^+_2,\quad \left\lbrace\bar{Q}_i,\bar{S}_j\right\rbrace = -\,m\,\varepsilon_{ij}\,{Z}^-_2,\nn
    &&\left[I^i_j,  I^k_l\right] = \delta^k_j\,I^i_l - \delta^i_l\,I^k_j\,,\quad
    \left[\bar{T},T\right]=2m\,\mathcal{H}_{\rm conf}\,,\quad\left[\mathcal{H}_{\rm conf}\,, \bar{T}\right]=-\,m\,\bar{T},\quad \left[\mathcal{H}_{\rm conf}\,, T\right]=m\,T,\nn
    &&\left[I^i_j, \bar{Q}_{l}\right] = \frac{1}{2}\,\delta^i_j\,\bar{Q}_{l}-\delta^i_l\,\bar{Q}_{j}\,,\qquad \left[I^i_j, Q^{k}\right]
    = \delta^k_j\,Q^{i} - \frac{1}{2}\,\delta^i_j\,Q^{k},\nn
    &&\left[I^i_j, \bar{S}_{l}\right] = \frac{1}{2}\,\delta^i_j\,\bar{S}_{l}-\delta^i_l\,\bar{S}_{j}\,,\quad \left[I^i_j, S^{k}\right]
    = \delta^k_j\,S^{i} - \frac{1}{2}\,\delta^i_j\,S^{k},\nn
        &&\left[\mathcal{H}_{\rm conf}\,, \bar{S}_{l}\right]=-\,\frac{m}{2}\,\bar{S}_{l}\,,\quad \left[\mathcal{H}_{\rm conf}\,, S^{k}\right]=\frac{m}{2}\,S^{k},\nn
        &&\left[\mathcal{H}_{\rm conf}\,, \bar{Q}_{l}\right]=\frac{m}{2}\,\bar{Q}_{l}\,,\quad
         \left[\mathcal{H}_{\rm conf}\,, Q^{k}\right]=-\,\frac{m}{2}\,Q^{k},\nn
        &&\left[T,Q^i\right]=-\,m\,S^i,\quad \left[T,\bar{S}_j\right]=-\,m\,\bar{Q}_j\,,\quad
        \left[\bar{T},\bar{Q}_j\right]=m\,\bar{S}_j\,,\quad \left[\bar{T},S^i\right]=m\,Q^i.
        \label{su211}
\eea
The superconformal algebra contains three central charges:
\bea
    Z_1 = -\,\cos{2\lambda}\;L, \qquad Z^{\pm}_2=\sin{2\lambda}\;L \pm ig.
\eea
Note that the quadratic operator constructed out of the central charges is reduced on the states to the square of the quantum number
$\tilde{l}=\sqrt{l^2 + g^2}$ defined in  \eqref{fermionicwf},
\bea
    \left(Z_1\right)^2 + Z^+_2Z^-_2= L^2 + g^2 := \tilde{L}^2. \label{square}
\eea
When $g=0\,$, this operator coincides with $L^2$. Below we will show that the set of three central charges in \eqref{su211} can be reduced
to a single central charge $\tilde{Z}_1 = \tilde{L}$\,. This agrees with the fact that the superalgebra $su(2|1,1)$ contains 15 generators:
eight supercharges, three $su(2)$ generators,
three generators of $so(2,1)$ and one central charge, in accord with the decomposition $su(2|1,1) = psu(2|1,1) \oplus \tilde{Z}_1$\,, where $psu(2|1,1)$ is a centerless superalgebra.

Before proceeding further, we point out that the SU$(2|1,1)$ trigonometric superconformal
model of the multiplet ${\bf (2,4,2)}$ with a superpotential term resulting in the Hamiltonian \eqref{confH} was constructed within a manifestly SU$(2|1)$ covariant (${\cal N}=4$ deformed)
superfield approach in \cite{ISTconf}. In the ${\cal N}=2, d=1$ superfield formalism, the model amounts to a system of the coupled ${\bf (2,2,0)}$ and ${\bf (0,2,2)}$ multiplets,
with only ${\cal N}=2$ superconformal symmetry SU$(1|1,1) \subset {\rm SU}(2|1,1)$ being manifest. In such a formulation,
the inverse-square terms with $g\neq 0$ in \eqref{confH}, \eqref{qH} come out solely from the coupling of these two multiplets
and disappear after decoupling of the fermionic multiplet ${\bf (0,2,2)}$. So in this limit (still respecting SU$(1|1,1)$ invariance) our model is reduced to
the two-dimensional ${\cal N}=2$ superconformal oscillator model based on the chiral multiplet ${\bf (2,2,0)}$.
In ref. \cite{CHT} (see also \cite{CT}) there was considered a different  SU$(1|1,1)$ superconformal model of the multiplet ${\bf (2,2,0)}$,
with the Hamiltonian involving an inverse-square potential induced by some spin coupling. It cannot be obtained as any truncation of our model.

\subsection{Casimir operators}\label{d21alpha}
Let us consider the quadratic Casimir operator of $D(2,1;\alpha)$ \cite{superc}:
\bea
    C^{\prime(\alpha)}_2 &=& \frac{2\mathcal{H}_{\rm conf}^2 - T\bar{T}-\bar{T}T}{2m^2}+ \frac{\alpha I^i_jI^j_i}{2}\,+\frac{Q^k\bar{Q}_k-\bar{Q}_kQ^k}{4m}
    -\frac{S^k\bar{S}_k-\bar{S}_kS^k}{4m}\nn
    &&-\,\frac{\left(\alpha+1\right)}{2}\left(F^2 + C\bar{C}\right).\label{C2primealpha}
\eea
The definition of the superalgebra $D(2,1;\alpha)$ in terms of these generators was given in \cite{ISTconf}, with $m=-\alpha\mu$.
The limit $\alpha = -1$ gives rise to the superalgebra $D(2,1;\alpha=-1)\rightarrow psu(2|1,1)\oplus su(2)\,$,
where $psu(2|1,1)$ is a centerless superalgebra and $su(2)$ is an external automorphism generated by the generators $F$, $C$ and $\bar{C}$.
Hence, the Casimir operator
of $psu(2|1,1)$ reads
\bea
    C^{\prime(\alpha=-1)}_2 = \frac{2\mathcal{H}_{\rm conf}^2- T\bar{T}-\bar{T}T}{2m^2} - \frac{I^i_jI^j_i}{2}+\frac{Q^k\bar{Q}_k-\bar{Q}_iQ^i}{4m}-\frac{S^k\bar{S}_k-\bar{S}_kS^k}{4m}.\label{C2prime}
\eea
The automorphism $su(2)$ generators $F$, $C$ and $\bar{C}$ commute with this operator.

In our case we deal with the centrally extended superalgebra \eqref{su211}, so we are led to perform an alternative way of reaching the limit $\alpha = -1$.
It implies the following preliminary redefinition of the extra $su(2)$ generators:
\bea
    Z_1=-\,2\left(\alpha+1\right)F,\qquad Z^+_2=-\,2\left(\alpha+1\right)C,\qquad Z^-_2=-\,2\left(\alpha+1\right)\bar{C}.\label{redef2}
\eea
Then, multiplying \eqref{C2primealpha} by the factor $\sim\left(\alpha+1\right)$ and taking the limit $\alpha=-1$ afterwards, we observe that only the last piece
of \eqref{C2primealpha} survives this limit:
\bea
    -\,8\left(\alpha+1\right)C^{\prime(\alpha)}_2 \quad\overset{\alpha=-1}{\Longrightarrow}\quad \left(Z_1\right)^2 + Z^+_2Z^-_2\,.
\eea
The generators \eqref{redef2} commute among themselves and with all other generators in the limit considered, so they form the triplet of central charges.
Thus we are left with the superalgebra \eqref{su211} for which the invariant operator \eqref{square} is a proper limit
of the quadratic Casimir operator \eqref{C2primealpha} and it is the genuine Casimir for the centrally extended $su(2|1,1)$ superalgebra.

Below we will show that SU$(2|1,1)$ is a spectrum-generating supersymmetry acting on infinite-dimensional irreducible SU$(2|1,1)$ multiplets
labeled by the eigenvalues $\tilde{l}^2$ of \eqref{square}. However, we must take into account that the angular momentum operator
$L$  can take two eigenvalues $l = \pm|l|$ leading to the same $(\tilde{l})^2$. So, the operator $L$ commuting with all $su(2|1,1)$ generators
can be treated as the second Casimir operator of $su(2|1,1)$ and the full space of quantum states of the model is the collection of two copies of SU$(2|1,1)$ multiplets,
with the same value of $\tilde{l}$ and two opposite-sign values of
the quantum number $l$.

\subsection{Passing to the new basis in SU$(2|1,1)$}
It is useful to bring the superconformal algebra \eqref{su211} to the form containing only one central charge.
To eliminate two of the original central charges,  we perform the rotation
\bea
    &&\tilde{Q}^i = \frac{\cos{\varphi}}{\sqrt{2}}\left(Q^i-i\bar{S}^i\right)e^{i\left(\lambda-\pi/4\right)} - \frac{\sin{\varphi}}{\sqrt{2}}\left(\bar{S}^i-iQ^i\right)e^{-i\left(\lambda-\pi/4\right)},\nn
    &&\bar{\tilde{Q}}_j = \frac{\cos{\varphi}}{\sqrt{2}}\left(\bar{Q}_j-iS_j\right)e^{-i\left(\lambda-\pi/4\right)} + \frac{\sin{\varphi}}{\sqrt{2}}\left(S_j-i\bar{Q}_j\right)e^{i\left(\lambda-\pi/4\right)},\nn
    &&\tilde{S}^i = \frac{\cos{\varphi}}{\sqrt{2}}\left(S^i-i\bar{Q}^i\right)e^{i\left(\lambda-\pi/4\right)} - \frac{\sin{\varphi}}{\sqrt{2}}\left(\bar{Q}^i-iS^i\right)e^{-i\left(\lambda-\pi/4\right)},\nn
    &&\bar{\tilde{S}}_j = \frac{\cos{\varphi}}{\sqrt{2}}\left(\bar{S}_j-iQ_j\right)e^{-i\left(\lambda-\pi/4\right)} + \frac{\sin{\varphi}}{\sqrt{2}}\left(Q_j-i\bar{S}_j\right)e^{i\left(\lambda-\pi/4\right)},\label{redef3}
\eea
where \footnote{Here, $\cos{2\varphi}$ and $\sin{2\varphi}$ are non-linear operators expanded as Taylor series over
the generator $L$. Since $L$ is a central charge, their series expansions take certain constant values.
In the case of an arbitrary parameter $\varphi$ such rotations amount to a particular ${\rm SO}(3)\sim {\rm SU}(2)$ external
group rotation preserving the invariant \eqref{square}.}
\bea
    \cos{2\varphi}=\frac{g}{\sqrt{L^2 + g^2}}\,,\qquad
    \sin{2\varphi}=\frac{L}{\sqrt{L^2 + g^2}}\,. \label{Defphi}
\eea
The only central charge $\tilde{Z}_1$ we are left with reads
\bea
    \tilde{Z}_1 = \sqrt{\left(Z_1\right)^2 + Z^+_2 Z^-_2} = \sqrt{L^2 + g^2}\,,
\eea
and it takes the value $\tilde{l} = \sqrt{l^2 + g^2}$ on the quantum states (see \eqref{square}).
In the new basis, the superalgebra is rewritten as
\bea
    &&\left\lbrace \tilde{Q}^i,\bar{\tilde{Q}}_j \right\rbrace = 2m\,I^i_j - m\,\delta^i_j\,\tilde{Z}_1 + 2\delta^i_j\,\mathcal{H}_{\rm conf}\,,\qquad
    \left\lbrace \tilde{S}^i,\bar{\tilde{S}}_j \right\rbrace = -\,2m\,I^i_j + m\,\delta^i_j\,\tilde{Z}_1 + 2\delta^i_j\,\mathcal{H}_{\rm conf}\,,\nn
    &&\left\lbrace \tilde{S}^i,\bar{\tilde{Q}}_j \right\rbrace = 2\delta^i_j\,T,\qquad \left\lbrace \tilde{Q}^i,\bar{\tilde{S}}_j \right\rbrace = 2\delta^i_j\,\bar{T},\nn
    &&\left[I^i_j,  I^k_l\right] = \delta^k_j\,I^i_l - \delta^i_l\,I^k_j\,,\quad
    \left[\bar{T},T\right]=2m\,\mathcal{H}_{\rm conf}\,,\quad\left[\mathcal{H}_{\rm conf}\,, \bar{T}\right]=-\,m\,\bar{T},\quad \left[\mathcal{H}_{\rm conf}\,, T\right]=m\,T,\nn
    &&\left[I^i_j, \bar{\tilde{Q}}_{l}\right] = \frac{1}{2}\,\delta^i_j\,\bar{\tilde{Q}}_{l}-\delta^i_l\,\bar{\tilde{Q}}_{j}\,,\qquad \left[I^i_j, \tilde{Q}^{k}\right]
    = \delta^k_j\,\tilde{Q}^{i} - \frac{1}{2}\,\delta^i_j\,\tilde{Q}^{k},\nn
    &&\left[I^i_j, \bar{\tilde{S}}_{l}\right] = \frac{1}{2}\,\delta^i_j\,\bar{\tilde{S}}_{l}-\delta^i_l\,\bar{\tilde{S}}_{j}\,,\qquad \left[I^i_j, \tilde{S}^{k}\right]
    = \delta^k_j\,\tilde{S}^{i} - \frac{1}{2}\,\delta^i_j\,\tilde{S}^{k},\nn
    &&\left[\mathcal{H}_{\rm conf}\,, \bar{\tilde{S}}_{l}\right]=-\,\frac{m}{2}\,\bar{\tilde{S}}_{l}\,,\quad \left[\mathcal{H}_{\rm conf}\,, \tilde{S}^{k}\right]=\frac{m}{2}\,\tilde{S}^{k},\nn
    &&\left[\mathcal{H}_{\rm conf}\,, \bar{\tilde{Q}}_{l}\right]=\frac{m}{2}\,\bar{\tilde{Q}}_{l}\,,\quad \left[\mathcal{H}_{\rm conf}\,, \tilde{Q}^{k}\right]=-\,\frac{m}{2}\,\tilde{Q}^{k},\nn
    &&\left[T,\tilde{Q}^i\right]=-\,m\,\tilde{S}^i,\quad \left[T,\bar{\tilde{S}}_j\right]=-\,m\,\bar{\tilde{Q}}_j\,,\quad
    \left[\bar{T},\bar{\tilde{Q}}_j\right]=m\,\bar{\tilde{S}}_j\,,\quad \left[\bar{T},\tilde{S}^i\right]=m\,\tilde{Q}^i.\label{su211m}
\eea

The generators $\tilde{Q}^i,\, \bar{\tilde{Q}}_j,\, I^i_j$ and
\be
\tilde{\mathcal{H}} = \mathcal{H}_{\rm conf} -\frac{m}{2}\,\tilde{Z}_1 \label{newH}
\ee
form a different $su(2|1)$ subalgebra, with the same (anti)commutators as in \eqref{su21}. So one can construct the new space of quantum SU$(2|1)$ states just with respect to
this transformed $su(2|1)$ superalgebra. It should be pointed out that the $su(2|1,1)$ generators in the original and new bases, taking into account their explicit form,
can be realized on the full set of the quantum states defined in Section 2, so that this set is closed under the action of these generators. Thereby, the construction
using the transformed $su(2|1)$ superalgebra will give rise to the same total set of quantum states, although with the energy spectrum calculated with respect to the Hamiltonian
$\tilde{\mathcal{H}}$ defined in \eqref{newH}.

The fermionic wave functions $\Psi^i_{(nl)}$ are still defined as in \eqref{fermionicwf}. Indeed, on the bosonic factor $\psi''\left(z,\bar{z}\right)$ the Hamiltonian
$\tilde{\mathcal{H}}$ can be represented as
\be
\tilde{\mathcal{H}} \rightarrow \mathcal{H} - \frac{m}{2}\left(\tilde{l} + l\cos{2\lambda}\right),
\ee
whence
\be
\tilde{\mathcal{H}}\,\Psi^i_{(n,l)} = \left[\mathcal{H} - \frac{m}{2}\left(\tilde{l} + l\cos{2\lambda}\right)\right] \Psi^i_{(nl)} = m \left( n - \frac12 \right)\Psi^i_{(n,l)}\,,\label{FermNew}
\ee
where we made use of the relation \eqref{Flevels}. Thus $\Psi^i_{(n,l)}$ are eigenfunctions of both $\mathcal{H}$ and $\tilde{\mathcal{H}}$.

The new definition of the bosonic wave functions is as follows
\bea
    &&\tilde{Q}^i\,\Psi^j_{(n,l)} =\left(n-1\right) \sqrt{2m}\,\varepsilon^{ij}\,\tilde{\Omega}^{-}_{(n,l)}\,,
\qquad \tilde{S}^i\,\Psi^j_{(n,l)} = \sqrt{2m}\,\varepsilon^{ij}\,\tilde{\Omega}^{-}_{(n+1,l)}\,,\nn
    &&\bar{\tilde{Q}}_j\,\Psi^i_{(n,l)} = \sqrt{2m}\,\delta^{i}_{j}\,\tilde{\Omega}^{+}_{(n,l)}\,,
\qquad \bar{\tilde{S}}_j\,\Psi^i_{(n,l)} =\left(n+\tilde{l}-1\right)\sqrt{2m}\,
\delta^{i}_{j}\,\tilde{\Omega}^{+}_{(n-1,l)}\,,\nn
    &&\tilde{Q}^i\,\tilde{\Omega}^{+}_{(n,l)}=n\,\sqrt{2m}\,\Psi^i_{(n,l)}\,,
\qquad \tilde{S}^i\,\tilde{\Omega}^{+}_{(n,l)}=\sqrt{2m}\,\Psi^i_{(n+1,l)}\,,\nn
    &&\bar{\tilde{Q}}^j\,\tilde{\Omega}^{-}_{(n,l)}=-\left(n-1\right)^2\sqrt{2m}\,\Psi^j_{(n,l)}\,,
\qquad \bar{\tilde{S}}^j\,\tilde{\Omega}^{-}_{(n,l)} = -\left(n + \tilde{l}-1\right)\sqrt{2m}\,\Psi^j_{(n-1,l)}\,,\nn
    &&\tilde{Q}^i\,\tilde{\Omega}^{-}_{(n,l)}=0,
\qquad \tilde{S}^i\,\tilde{\Omega}^{-}_{(n,l)}=0,\qquad
    \bar{\tilde{Q}}^j\,\tilde{\Omega}^{+}_{(n,l)}=0,
\qquad \bar{\tilde{S}}^j\,\tilde{\Omega}^{+}_{(n,l)} = 0,
    \label{QPsi}
\eea
where
\bea
    n=0, 1, 2 \ldots \quad {\rm for} \quad \tilde{\Omega}^{+}_{(n,l)}\,, \qquad {\rm and} \qquad
    n=2, 3, 4 \ldots \quad {\rm for} \quad \tilde{\Omega}^{-}_{(n,l)}\,.
\eea
They can be expressed through the wave functions \eqref{omega12} and \eqref{omega0},
\bea
    \tilde{\Omega}^{-}_{(n,l)}=\frac{\left(\cos{\varphi}+\sin{\varphi}\right)e^{-i\pi /4}}{\sqrt{2}\left(\tilde{l}+l\right)}\,\Omega^{2}_{(n-1,l)}\,,\qquad
    \tilde{\Omega}^{+}_{(n,l)}=\frac{\sqrt{2}\,\cos{\varphi}\;e^{-i\pi /4}}{\tilde{l}\left(\tilde{l}+g-l\right)}\,\Omega^{1}_{(n,l)}\,,\label{tildeomega}
\eea
and, further, through ${\Omega}^{\pm}_{(n,l)}$\,. The conformal generators $T$ and $\bar{T}$ are realized on the new wave functions
as the creation and annihilation operators:
\bea
    &&T\,\tilde{\Omega}^{-}_{(n,l)}=m\,\tilde{\Omega}^{-}_{(n+1,l)}\,,\qquad \bar{T}\,\tilde{\Omega}^{-}_{(n,l)}=\left(n-2\right)\left(n+\tilde{l}-1\right)m\,\tilde{\Omega}^{-}_{(n-1,l)}\,,\nn
    &&T\,\Psi^i_{(n,l)}=m\,\Psi^i_{(n+1,l)}\,,\qquad \bar{T}\,\Psi^i_{(n,l)}=\left(n-1\right)\left(n+\tilde{l}-1\right)m\,\Psi^i_{(n-1,l)}\,,\nn
    &&T\,\tilde{\Omega}^{+}_{(n,l)}=m\,\tilde{\Omega}^{+}_{(n+1,l)}\,,\qquad \bar{T}\,\tilde{\Omega}^{+}_{(n,l)}=n\left(n+\tilde{l}-1\right)m\,\tilde{\Omega}^{+}_{(n-1,l)}\,.
\eea

Now, the full energy spectrum of the Hamiltonian $\tilde{\mathcal{H}} = \mathcal{H}_{\rm conf}-m\,\tilde{Z}_1/2$ can be given
by the simple single formula as the spectra of three harmonic oscillators with the same frequency,
\bea
    E^{\sigma}_{(n)}=m\left(n-\sigma\right), \label{En}
\eea
where
\bea
    &&\sigma=0,\quad n=0, 1, 2 \ldots \quad {\rm for} \quad \tilde{\Omega}^{+}_{(n,l)}\,,\nn
    &&\sigma=1/2,\quad n=1, 2, 3 \ldots \quad {\rm for} \quad \Psi^{+}_{(n,l)}\,,\nn
    &&\sigma=1,\quad n=2, 3, 4 \ldots \quad {\rm for} \quad \tilde{\Omega}^{-}_{(n,l)}
\eea
[cf. \eqref{FermNew}]. Thus, the bosonic and fermionic states occupy integer and half integer energy values, respectively.
In other words, they fill even and odd levels.  

The new SU$(2|1)$ supersymmetry generated by the supercharges $\tilde{Q}^i$ and $\bar{\tilde{Q}}_j$
is not spontaneously broken and the corresponding ground state is given by the SU$(2|1)$ singlet $\tilde{\Omega}^{+}_{(0,l)}$\,.
To prove this, let us consider the relevant Casimir operators
\bea
    \tilde{C}_2 &=& \frac{1}{m^2}\left(\mathcal{H}_{\rm conf}-\frac{m}{2}\,\tilde{Z}_1\right)^2-\frac{I^i_jI^j_i}{2}+\frac{\tilde{Q}^i\bar{\tilde{Q}}_i-\bar{\tilde{Q}}_i\tilde{Q}^i}{4m}\,,\nn
    \tilde{C}_3 &=& \frac{1}{m}\left(\tilde{C}_2+\frac{1}{2}\right)\left(\mathcal{H}_{\rm conf}-\frac{m}{2}\,\tilde{Z}_1\right)\nn
    &&+\,\frac{1}{8m^2}\left(\delta^i_j \mathcal{H}_{\rm conf}-\frac{m}{2}\,\delta^i_j\tilde{Z}_1-m\,I^i_j\right)\left(\tilde{Q}^j\bar{\tilde{Q}}_i
    -\bar{\tilde{Q}}_i\tilde{Q}^j\right).\label{tildeC23}
\eea
On the states $\tilde{\Omega}^{\pm}_{(n,l)}$ and $\Psi^i_{(n,l)}$\,, they take the eigenvalues
\bea
    \tilde{C}_2 = n\left(n - 1\right),\qquad \tilde{C}_3 = \left(n - \frac{1}{2}\right)\tilde{C}_2\,.\label{valuestildeC23}
\eea
The state $\tilde{\Omega}^{+}_{(0,l)}$ thus corresponds to zero eigenvalues of both Casimir operators:
\bea
    \tilde{C}_2\,\tilde{\Omega}^{+}_{(0,l)} = 0,\qquad  \tilde{C}_3\,\tilde{\Omega}^{+}_{(0,l)} = 0.
\eea
It follows from the relations \eqref{QPsi} that $\tilde{Q}^i\tilde{\Omega}^{+}_{(0,l)} = \bar{\tilde{Q}}_j\tilde{\Omega}^{+}_{(0,l)}=0$, as it should be for the singlet
ground state.

The states with $n = 1$ correspond to an atypical representation, since Casimir operators are also zero on these states.
This representation is spanned by the three states
\bea
    \Psi^i_{(1,l)}\,,\qquad \bar{\tilde{Q}}_j\,\Psi^i_{(1,l)} = \sqrt{2m}\,\delta^{i}_{j}\,\tilde{\Omega}^{+}_{(1,l)}\,,
\eea
such that
$$
\tilde{Q}^j\Psi^i_{(1,l)} = 0.
$$
The states $ \left\lbrace\Psi^i_{(1,l)}, \tilde{\Omega}^{+}_{(1,l)}\right\rbrace$ form the fundamental SU$(2|1)$ representation.
All excited states with $n > 1$ correspond to the simplest four-fold typical SU$(2|1)$ representations.

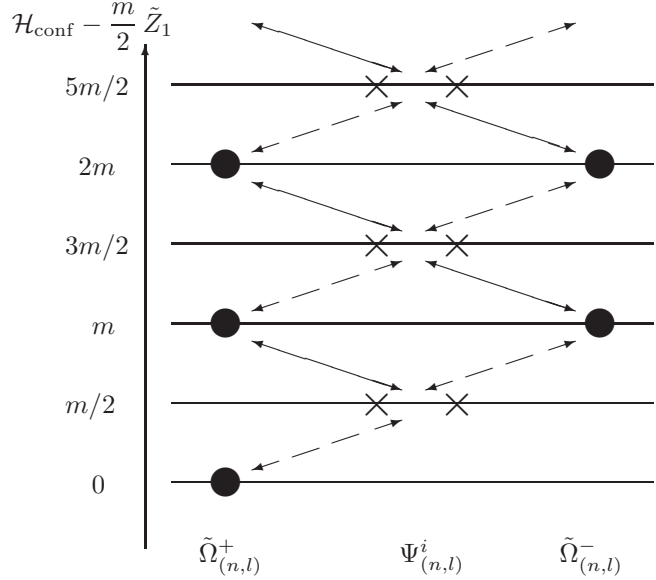
\begin{figure}[h]
\begin{center}
\begin{picture}(240,210)

\put(50,5){\line(0,1){190}}
\put(50,185){\vector(0,1){10}}

\put(60,30){\line(1,0){180}}
\put(60,90){\line(1,0){180}}
\put(60,150){\line(1,0){180}}
\put(60,60){\line(1,0){180}}
\put(60,120){\line(1,0){180}}
\put(60,180){\line(1,0){180}}

\put(80,30){\circle*{11}}
\put(80,90){\circle*{11}}
\put(80,150){\circle*{11}}

\multiput(130,55)(30,0){2}{{\LARGE $\times$}}
\multiput(130,115)(30,0){2}{{\LARGE $\times$}}
\multiput(130,175)(30,0){2}{{\LARGE $\times$}}

\put(220,90){\circle*{11}}
\put(220,150){\circle*{11}}

\put(30,26){$0$}
\put(20,56){$m/2$}
\put(30,86){$m$}
\put(20,116){$3m/2$}
\put(25,146){$2m$}
\put(20,176){$5m/2$}
\put(0,200){$\displaystyle \mathcal{H}_{\rm conf}-\frac{m}{2}\,\tilde{Z}_1$}
\put(145,0){$\Psi^i_{(n,l)}$}
\put(70,0){$\tilde{\Omega}^+_{(n,l)}$}
\put(205,0){$\tilde{\Omega}^-_{(n,l)}$}

\put(100,80){\line(3,-1){45}}
\put(100,80){\vector(-3,1){10}}
\put(136,68){\vector(3,-1){10}}

\put(100,140){\line(3,-1){45}}
\put(136,128){\vector(3,-1){10}}
\put(100,140){\vector(-3,1){10}}

\put(100,200){\line(3,-1){45}}
\put(100,200){\vector(-3,1){10}}
\put(136,188){\vector(3,-1){10}}

\put(165,110){\vector(-3,1){10}}
\put(165,110){\line(3,-1){45}}
\put(201,98){\vector(3,-1){10}}

\put(165,170){\vector(-3,1){10}}
\put(165,170){\line(3,-1){40}}
\put(201,158){\vector(3,-1){10}}

\put(130,48){\line(-3,-1){10}}
\put(115,43){\line(-3,-1){10}}
\put(100,38){\vector(-3,-1){10}}
\put(136,50){\vector(3,1){10}}

\put(195,78){\line(-3,-1){10}}
\put(180,73){\line(-3,-1){10}}
\put(165,68){\vector(-3,-1){10}}
\put(201,80){\vector(3,1){10}}

\put(195,138){\line(-3,-1){10}}
\put(180,133){\line(-3,-1){10}}
\put(165,128){\vector(-3,-1){10}}
\put(201,140){\vector(3,1){10}}

\put(195,198){\line(-3,-1){10}}
\put(180,193){\line(-3,-1){10}}
\put(165,188){\vector(-3,-1){10}}
\put(201,200){\vector(3,1){10}}

\put(130,108){\line(-3,-1){10}}
\put(115,103){\line(-3,-1){10}}
\put(100,98){\vector(-3,-1){10}}
\put(136,110){\vector(3,1){10}}

\put(136,170){\vector(3,1){10}}
\put(130,168){\line(-3,-1){10}}
\put(115,163){\line(-3,-1){10}}
\put(100,158){\vector(-3,-1){10}}

\end{picture}
\end{center}
\caption{Degeneracy of energy levels with a fixed number $l$. The action of the supercharges $\tilde{Q}^i$ and $\bar{\tilde{Q}}_j$
is drawn by solid lines, while dashed lines corresponds to the action of $\tilde{S}^i$ and $\bar{\tilde{S}}_j$\,.}
\label{figure2}
\end{figure}

Whereas the supersymmetry associated with $\tilde{Q}^i$ and $\bar{\tilde{Q}}_j$ is not broken, the second pair of SU$(2|1)$ supercharges $\tilde{S}^i$ and $\bar{\tilde{S}}_j$ corresponds
to the spontaneously broken supersymmetry (see Figure \ref{figure2}),
with the minimal energy $\tilde{l}\,m$
as the lowest eigenvalue of the relevant shifted Hamiltonian $\mathcal{H}_{\rm conf}+m\,\tilde{Z}_1/2$.

On Figure \ref{figure2}, we demonstrate how an action of SU$(2|1,1)$ supercharges mixes all bosonic and fermionic states with a fixed number of the angular momentum $l$.
Thus, all these states at fixed $l$ belong to a single infinite-dimensional SU$(2|1,1)$ representation labeled by $\tilde{l}^2=l^2+g^2$ as the square of the central charge.
The states with $-\,l$ belong to the other SU$(2|1,1)$ representation labeled by the same $\tilde{l}^2=l^2+g^2$.
These two representations are distinguished only by the eigenvalue of the operator $L$ (see discussion in Subsection \ref{d21alpha}) and they exhaust
the whole space of the quantum states of the model. Thus the conformal supergroup $su(2|1,1)$ acts as the spectrum-generating algebra on this space
of quantum states\footnote{The analogous role of the superconformal algebras $D(2,1;\alpha)$ and
$su(1|1,1)$ in the quantum Hilbert spaces of some superconformal mechanics models was pointed out in \cite{CHT}, \cite{FIS}, \cite{CT}.}.

Taking into account \eqref{QPsi}, we can define the generators
\bea
    B_+=\frac{1}{2m}\,\tilde{S}_k\tilde{Q}^k,\qquad B_-=\frac{1}{2m}\,\bar{\tilde{Q}}_k\bar{\tilde{S}}^k, \label{Bpm}
\eea
which are responsible for the two-fold degeneracy of exited bosonic states:
\bea
    &&B_+\,\tilde{\Omega}^{+}_{(n,l)} = 2\,\tilde{\Omega}^{-}_{(n+1,l)}\,,\qquad n=1,2,3, \ldots , \nn
    &&B_-\,\tilde{\Omega}^{-}_{(n+1,l)}=-\,2n\left(\tilde{l}+n\right)\tilde{\Omega}^{+}_{(n,l)}\,,\quad n=1,2,3 \ldots .
\eea
These generators act on the bosonic wave functions only and form an exotic SU(2) symmetry \cite{DSQM} belonging to the universal enveloping of $su(2|1,1)$:
\bea
    &&\left[B_+, B_-\right]=2B_3\,,\qquad  \left[B_3, B_\pm\right]=\mp\,8n\left(\tilde{l}+n\right)B_\pm\,, \nn
    &&B_3\,\tilde{\Omega}^{+}_{(n,l)}=2n\left(\tilde{l}+n\right)\tilde{\Omega}^{+}_{(n,l)}\,,\quad
    B_3\,\tilde{\Omega}^{-}_{(n+1,l)}=-\,2n\left(\tilde{l}+n\right)\tilde{\Omega}^{-}_{(n+1,l)}\,,\quad n=1,2,3, \ldots .\qquad
\eea
The ground state $\tilde{\Omega}^{+}_{(0,l)}$ is annihilated by these SU(2) generators.
This SU(2) symmetry is responsible as well for the double-fold degeneracy of the initial wave functions $\Omega^{\pm}_{(n,l)}$\,, since the latter
are related to $\tilde{\Omega}^{\pm}_{(n,l)}$ via \eqref{omega-omega12} and \eqref{tildeomega}.

\subsection{Invariant operators}
The Casimir operator \eqref{C2prime} allows us to guess the form of some new invariant operators $\tilde{I}$, $\tilde{M}$
of the $su(2|1)$ superalgebra. They commute with $\tilde{Q}^i$, $\bar{\tilde{Q}}_j$\,, $I^i_j$\,, $\tilde{Z}_1$\,,
$\mathcal{H}_{\rm conf}$ and read \footnote{The operator $\tilde{I}$ becomes the quadratic Casimir \eqref{C2prime} of $psu(2|1,1)$
in the limit $\tilde{Z}_1=0$, where $\tilde{Q}^i\equiv Q^i$.}
\bea
    &&\tilde{I}=\frac{2\mathcal{H}_{\rm conf}^2- T\bar{T}-\bar{T}T}{2m^2}-\frac{I^i_jI^j_i}{2}
+\frac{\tilde{Q}^k\bar{\tilde{Q}}_k-\bar{\tilde{Q}}_k\tilde{Q}^k}{4m}
-\frac{\tilde{S}^k\bar{\tilde{S}}_k-\bar{\tilde{S}}_k\tilde{S}^k}{4m}
- \frac{\mathcal{H}_{\rm conf}\,\tilde{Z}_1}{m}+\frac{\left(\tilde{Z}_1\right)^2}{4}\,,\nn
    &&\tilde{M}=\frac{T\bar{T}}{2m^2}+\frac{\tilde{S}^k\bar{\tilde{S}}_k}{4m}\,.\label{inv}
\eea
On the SU$(2|1)$ representations generated by $\tilde{Q}^i$ and $\bar{\tilde{Q}}_j$ they take the values
\bea
    \tilde{I} = \tilde{l}\left(\frac{1}{2}-n\right),\qquad \tilde{M}=\frac{n}{2}\left(n+\tilde{l}-1\right).
\eea
The quadratic SU$(2|1)$ Casimir can be written in terms of these operators and the central charge:
\bea
    \tilde{C}_2= \tilde{I}+2\tilde{M}-\frac{\tilde{Z}_1}{2}\,.
\eea
These expressions will help us to construct generalizations of Uhlenbeck tensor in the next Section. Note that such invariants
can be constructed only for $su(2|1)$ generated by the transformed supercharges $\tilde{Q}^i$ and $\bar{\tilde{Q}}_j$\,. No their
analogs commuting with ${Q}^i$ and $\bar{Q}_j$ exist.

\subsection{Summary of Section 3}
The superconformal Hamiltonian $\mathcal{H}_{\rm conf}$\,, eq. \eqref{confH}, differs from the original SU($2|1)$ Hamiltonian $\mathcal{H}$, eq. \eqref{qH2},
by a central charge generator, eq.  \eqref{redef}.
The whole set of the superconformal SU($2|1,1$) generators, including $\mathcal{H}_{\rm conf}$\,, is realized on the original Hilbert space. The latter can be equally restored,
starting from a new $su(2|1) \subset su(2|1,1)$ which is related to the original one by the transformation \eqref{redef3}, \eqref{Defphi} and involves the
Hamiltonian $\tilde{\mathcal{H}} = \mathcal{H}_{\rm conf}-m\,\tilde{Z}_1/2$. The new ground state proves to be SU($2|1)$ singlet and possesses zero energy, so the redefined SU($2|1)$
symmetry is not spontaneously broken, in contrast to the original one. The fermionic wave functions are the same as in the original SU($2|1)$ model, while the bosonic ones
$\tilde{\Omega}^{\pm}$, including the new ground state, are represented by the proper linear combinations of the original wave functions $\Omega^{\pm}$.
In the new basis the dependence of the spectrum \eqref{En} and Casimir's eigenvalues \eqref{valuestildeC23} on the original parameter $\lambda$
vanishes. This phenomenon is due to superconformal symmetry and generalizes to the quantum case the property observed in \cite{ISTconf} at the classical level.

To avoid a possible confusion, we point out that the complete  quantum consideration of the SU$(2|1)$ supersymmetric $\mathbb{C}^N$ S.-W. system,
including energy spectrum, the structure
of the Hilbert space of wave functions and their SU$(2|1)$ representation contents, has been already given in Section 2. The basic aim of Section 3 was to demonstrate that the same
 results can be restored, starting from an equivalent description of this model in terms of complex SU$(2|1,1)$ superconformal mechanics associated with the
 supermultiplet ${\bf (2, 4, 2)}$. Many peculiar features of the original formulation become simpler in the superconformal formulation, including, e.g., simplifying
the formula for the energy spectrum. The phenomenon of disappearance of the dependence on the parameter $\lambda$ in the second formulation also deserves an attention.

\section{Supersymmetric $\mathbb{C}^N$ Smorodinsky--Winternitz system}
We define the quantum SU$(2|1)$ supersymmetric $\mathbb{C}^N$ S.-W. system as a sum of $N$ copies
of the $\mathbb{C}^1$ system with the Hamiltonians \eqref{qH} involving the same parameters $B$, $\omega$ (equivalently, $m$, $\lambda$),
\be
\mathcal{H}=\sum_{a=1}^N\mathcal{H}_a\,, \qquad
 \mathcal{H}_a= \bar{\pi}_a\pi_a+\omega^2 z^a\bar{z}^a+\frac{\left(g_a\right)^2}{4 z^a\bar z^a}
 +ig_a\left[\frac{\xi^a_k\xi^{ka}}{4\left(z^a\right)^2}-\frac{\bar{\xi}^{ka}\bar{\xi}^a_k}{4\left(\bar{z}^a\right)^2}\right]+\frac{B}{2}\,\xi^{ak}\bar\xi^a_k\,.
\label{SWsuper}
\ee
The supercharges and R-charges which form, together with the Hamiltonian $\mathcal{H}$, $su(2|1)$ superalgebra,
are also defined as sums of the relevant quantities of each particular $\mathbb{C}^1$-system.
Clearly, the generators $\mathcal{H}_a$ commute with each other, and thus define the constants of motion
of the supersymmetric $\mathbb{C}^{N}$ S.-W. system.
In addition to $N$ commuting integrals $\mathcal{H}_a$\,, this system possesses $N$ manifest U(1) symmetries  $z^a\to {\rm e}^{i\kappa}z^a, \;\xi^a_i\to {\rm e}^{i\kappa}\xi^a_i$,
with the generators
\be
{L}_{a}=z^a\partial_a -\bar z^a\partial_{\bar a}+\xi^{ak}\bar\xi^a_k- 1\,:\quad \left[{L}_{a}, \,{L}_{b}\right]= \left[{L}_{a}, \,\mathcal{H}_{b}\right]=0\,.
\label{sLa}
\ee
Hence, these generators provide the system to be integrable.

The wave functions of this system are obviously given by the products of those of $N$ one-dimensional copies,
and the energy spectrum -- by the sum of energies \eqref{Flevels1},
\bea
    E^{\sigma}_{(n , l)}=\sqrt{4\omega^2 +B^2}\left(n-\sigma+\frac{\tilde{l}}{2}\right)+ \frac{Bl}{2}\,,
\eea
where
\bea
    n=\sum_{a=1}^N n_a\,,\qquad l=\sum_{a=1}^N l_a\,,\qquad \tilde{l}=\sum_{a=1}^N\tilde{l}_a =\sum_{a=1}^N \sqrt{l^2_a+g^2_a}\,,\qquad \sigma=\sum_{a=1}^N\sigma_a\,.\label{numbers}
\eea
One observes the same distinction  between the spectra of bosonic and fermionic wave functions as in the $\mathbb{C}^1$ case.

The SU$(2|1)$ supersymmetric $\mathbb{C}^N$ S.-W. system has an additional degeneracy of the spectrum.
It is due to the existence of the additional constants of motion given by the components of the supersymmetric extension of
the Uhlenbeck tensor generating a hidden symmetry in the bosonic case.
The classical version of this supersymmetric Uhlenbeck tensor was constructed in \cite{ClSW}, while
its quantum counterpart can be written in the form
\bea
    I_{ab} &=&\frac{1}{4}\left(z^a\partial_a + \bar{z}^a\partial_{\bar a}+1\right)\left(z^b\partial_b
    + \bar{z}^b\partial_{\bar b}+1\right)-\frac{1}{2}\left(z^a\bar{z}^a\,\partial_b\partial_{\bar b} + z^b\bar{z}^b\,\partial_a\partial_{\bar a}\right)\nn
     &&+ \frac{i}{8}\left[g_b\Big(\frac{z^a\bar{z}^a}{z^bz^b}\,\xi_k^{b}\xi^{kb}-\frac{z^a\bar{z}^a}{\bar{z}^b\bar{z}^b}\,\bar{\xi}^k_{b}\bar{\xi}_{kb}\Big)+
     g_a\Big(\frac{z^b\bar{z}^b}{z^az^a}\,\xi_k^{a}\xi^{ka}-\frac{z^b\bar{z}^b}{\bar{z}^a\bar{z}^a}\,\bar{\xi}^k_{a}\bar{\xi}_{ka}\Big)\right]\nn
     &&+\,\frac{\left(g_b\right)^2z^a\bar{z}^a}{8z^b\bar{z}^b}+\frac{\left(g_a\right)^2z^b\bar{z}^b}{8z^a\bar{z}^a}
     -\,\frac{\delta_{ab}}{2}\left(z^a\partial_a + \bar{z}^a\partial_{\bar a}+1\right),
\label{sUh}
\eea
where no sum over $a$ and $b$ is assumed. These constants of motion, together with  \eqref{sLa} and $\mathcal{H}_a\,$, provide
the system with the superintegrability property.

 It turns out that this tensor admits a convenient representation in terms of the generators of the
associated superconformal algebra $su(2|1,1)$.

\subsection{Superconformal view}
Let us define the superconformal Hamiltonian  on $\mathbb{C}^N$ as a sum of
$N$ copies of superconformal Hamiltonians on $\mathbb{C}^1$  ,
\bea
    \mathcal{H}_{\rm conf} = \sum_{a=1}^{N}\mathcal{H}_{({\rm conf})\; a}\,,\label{confHsum}
\eea
where $\mathcal{H}_{({\rm conf})\; a}$ is given by \eqref{confH}, with different parameters $g_a$ for each $a$ ($a=1 \ldots N$),
but with the common parameters $\lambda$ and $m$\,. So we deal with a direct sum of $su(2|1,1)$ algebras labeled by the index $a$. Then, we take sums of all these generators
\bea
    &&\tilde{Q}^{i} = \sum_{a=1}^{N}\left(\tilde{Q}_a\right)^{i},\qquad \tilde{S}^{i} = \sum_{a=1}^{N}\left(\tilde{S}_a\right)^{i},\qquad
    \bar{\tilde{Q}}_{j} = \sum_{a=1}^{N}\left(\bar{\tilde{Q}}_{a}\right)_{j}\,,\qquad \bar{\tilde{S}}_{j} = \sum_{a=1}^{N}\left(\bar{\tilde{S}}_{a}\right)_{j}\,,\nn
    &&I^{i}_{j} = \sum_{a=1}^{N}\left(I_a\right)^{i}_{j}\,,\qquad T = \sum_{a=1}^{N}T_a\,,\qquad \bar{T} = \sum_{a=1}^{N}\bar{T}_a\,,\qquad
    \tilde{Z}_1 = \sum_{a=1}^{N}\tilde{Z}_{1a}\,,\label{sum}
\eea
and obtain, once again, the conformal superalgebra \eqref{su211m} with the superconformal Hamiltonian \eqref{confHsum}.
Here, $\tilde{Z}_{1a}$ are defined as
\bea
    \tilde{Z}_{1a}=\sqrt{\left(L_a\right)^2 + g_a^2}\,.
\eea

The wave eigenfunctions of \eqref{confHsum} are obviously the products of $N$ wave functions corresponding to $a=1 \ldots N$.
Taking into account \eqref{numbers}, the energy spectrum of the SU$(2|1)$ Hamiltonian $\tilde{H} = \mathcal{H}_{\rm conf}-m\,\tilde{Z}_1/2$ is given
by the obvious generalization of the formula \eqref{En}
\bea
    E^{\sigma}_{(n)}=m\left(n-\sigma\right), \label{EnN}
\eea
where $\sigma$ is a sum of $\sigma_a=0,1/2,1$ and it takes integer and half integer values ranged from $0$ to $N$. Bosonic and fermionic states
still occupy separate levels with integer and half-integer values of the energy (modulo the overall parameter $m$), respectively.

The Uhlenbeck tensor \eqref{sUh} commutes with the superconformal Hamiltonian \eqref{confHsum}. In terms of the generators of the conformal algebra $so(2,1)$
it can be represented in the very simple form
\bea
    I_{ab} = \frac{1}{2}\left[H_{({\rm conf})\; a}\,K_b+K_a\,H_{({\rm conf})\; b}\right] - D_aD_b\,,\label{Iab0}
\eea
or, in the basis \eqref{HKD},
\bea
    I_{ab} &=& \frac{1}{m^2}\left[\mathcal{H}_{({\rm conf})\; a}\,\mathcal{H}_{({\rm conf})\; b}-\frac{1}{2}\left(T_{a}\bar{T}_{b}+\bar{T}_{a}T_{b}\right)\right] \nonumber \\
    &=& \frac{1}{m^2}\left[\mathcal{H}_{({\rm conf})\; a}\,\mathcal{H}_{({\rm conf})\; b} - m\, \delta_{ab}\,\mathcal{H}_{({\rm conf})\; a}\right]
    - \big(M_{a\bar{b}} + M_{b\bar{a}}\big),  \label{Iab}
\eea
where
\bea
    M_{a\bar b}:=\frac{1}{2m^2}\,T_{a}\bar{T}_{b}\,.\label{Mab}
\eea
The second form of $I_{ab}$ makes obvious its commutativity with the superconformal Hamiltonian \eqref{confHsum}, as
well as with the Hamiltonian of $\mathbb{C}^N$ S.-W. system
\eqref{SWsuper}.

The non-linear algebra generated by $I_{ab}$ reads
\bea
    \left[I_{ab},I_{cd}\right]&=& \delta_{ac}\,T_{cbd} + \delta_{ad}\,T_{dbc} + \delta_{bc}\,T_{cad} + \delta_{bd}\,T_{dac}\,,
  \qquad a\neq b,\; c\neq d\,,\nn
 \left[I_{aa},I_{cd}\right]&=&0 \label{nlalgebra0}
\eea
(no summation over repeated indices), where the function $T_{cbd}$ has a simple representation  through the generators of conformal algebra:
\bea
    T_{cbd}=\frac{1}{m}\,\big[\mathcal{H}_{({\rm conf})\; c}\,\big(M_{b\bar{d}} - M_{d\bar{b}}\big) + \mathcal{H}_{({\rm conf})\; d}\,\big(M_{c\bar{b}} - M_{b\bar{c}}\big)
    +\mathcal{H}_{({\rm conf})\; b}\,\big(M_{d\bar{c}} - M_{c\bar{d}}\big)\big].
     \label{nlalgebra}
\eea
Notice that for calculating the commutation relations \eqref{nlalgebra0} we do not need the explicit expressions for $I_{ab}$ in terms of the variables
$(z^a, {\bar z}^a, \xi^{ai}, \bar\xi^{a}_{ i})$ as in \eqref{sUh},  now it suffices to make use of
the standard commutation relations \eqref{so210} or \eqref{so21} of  the conformal algebra $so(2,1)\,$.

Looking at the expressions \eqref{Iab} and \eqref{nlalgebra}, we observe that they involve, apart from $N$ Hamiltonians $\mathcal{H}_{({\rm conf})a}$,
also $N^2$ bilinear generators $M_{a\bar{b}}$ that commute with \eqref{confHsum} and \eqref{SWsuper}.
Thus, what actually matters is the nonlinear closed  algebra generated by $M_{a\bar{b}}$ and $\mathcal{H}_{({\rm conf})\; b}$:
\bea
    &&\left[M_{a\bar b}, \, M_{b\bar c}\right]=\frac{1}{m}\,M_{a\bar{c}}\,\mathcal{H}_{({\rm conf})\; b}\,,\qquad a\neq b,\;\; b\neq c,\;\; c\neq a,\nn
    &&\left[M_{b\bar b}, \, M_{b\bar c}\right]=\frac{1}{m}\,M_{b\bar{c}}\,\mathcal{H}_{({\rm conf})\; b}\,,\qquad b\neq c, \nn
    && \left[ M_{a\bar b}, \, M_{b \bar b} \right] = \frac{1}{m} M_{a\bar b} \Big(\mathcal{H}_{({\rm conf})\;b}- m\Big), \qquad  b\neq a,  \label{MDef} \\
    &&  \left[\mathcal{H}_{({\rm conf})\; b}, \, M_{c\bar{d}}\right] = m\Big(\delta_{bc}- \delta_{bd}\Big) M_{c\bar{d}} \label{MHDef}
\eea
(no summation over indices). One can add to this set the U(1) generators $L_a\,$, which commute with everything. Note that the symmetric combination $M_{a\bar{b}} + M_{b\bar{a}}$
can be directly expressed through $\mathcal{H}_{({\rm conf})\; b}$ and $I_{ab}$ from \eqref{Iab}, but it is not true for the antisymmetric one $M_{a\bar{b}} - M_{b\bar{a}}$ entering
$T_{abc}$. However, it is possible to express $(M_{a\bar{b}} - M_{b\bar{a}})^2$ through the rest of constants of motion:
\bea
&&\left(M_{a\bar{b}} - M_{b\bar{a}}\right)^2 = \left(M_{a\bar{b}} + M_{b\bar{a}}\right)^2 - 4 M_{a\bar{a}}M_{b\bar{b}}\,, \quad a\neq b\,, \label{MinPlus} \\
&& M_{a\bar{b}} + M_{b\bar{a}} = \frac1{m^2}\,\mathcal{H}_{({\rm conf})\; a}\mathcal{H}_{({\rm conf})\; b} -I_{ab}\,,\quad a\neq b\,,\label{Plus} \\
&&2M_{a\bar{a}} = \frac1{m^2}\left[\mathcal{H}_{({\rm conf})\; a}\,\mathcal{H}_{({\rm conf})\; a} - m\,\mathcal{H}_{({\rm conf})\; a}\right] -I_{aa}\,. \label{Maa}
\eea
Thus the quantity $T_{cbd}$ defined in \eqref{nlalgebra} is a function of the original hidden symmetry generators $\mathcal{H}_{({\rm conf})\; a}, I_{cd}$
and so the relations \eqref{nlalgebra0}, \eqref{nlalgebra} constitute a closed non-linear algebra which is equivalent to the algebra \eqref{MDef}, \eqref{MHDef}.

Like in the bosonic $\mathbb{C}^1$ model (see \eqref{B10}), the diagonal integrals $I_{aa}$ are yet expressed through other integrals:
\bea
&& I_{aa}  = \frac14 \left(L_a^2 + g_a^2 \right)- \frac{I_a^2}{6} + A_a\,,\label{Square} \\
&& I_a^2 := \left(I_a\right)^{i}_{j} \left(I_a\right)^{j}_{i}\,, \qquad  A_a := \frac{ig_a}{4}
\left(\frac{\bar{z}^a}{z^a}\,\xi^a_k\xi^{ak} - \frac{{z}^a}{\bar{z}^a}\, \bar{\xi}^{ak}\bar{\xi}^{a}_k\right) - \frac12\left(\xi^{ak}\bar\xi^a_k - 1\right)L_a\,, \label{NewDef}
\eea
with
\bea
\left[\mathcal{H}_{({\rm conf})\; a}\,, I^2_a\right] = \left[\mathcal{H}_{({\rm conf})\; a}\,, A_a\right] = 0\,, \quad
A_a^2 = \frac14\left(L_a^2 + g_a^2\right)\left(1 - \frac{2I_a^2}{3}\right). \label{HAF}
\eea
The operators  $I_a^2$ are Casimirs for $N$ copies of SU(2) symmetries acting only on fermionic variables. The additional new integrals  of motion
$A_a$ can be written in terms of the superconformal SU$(2|1,1)$ generators as
\bea
    A_a =\frac{I_a^2}{3}+\frac{1}{8m}\left[\left(\tilde{S}_a\right)^k\left(\bar{\tilde{S}}_{a}\right)_k-\left(\bar{\tilde{S}}_{a}\right)_k\left(\tilde{S}_a\right)^k\right]
    -\frac{1}{8m}\left[\left(\tilde{Q}_a\right)^k\left(\bar{\tilde{Q}}_{a}\right)_k-\left(\bar{\tilde{Q}}_{a}\right)_k\left(\tilde{Q}_a\right)^k\right]. \label{Aa}
    \eea
In Appendix A we adduce some further details on the structure of these extra constants of motion.

Degeneracy of the energy spectrum \eqref{EnN} can be attributed to any operator commuting with the Hamiltonian.
One can construct many examples of such operators like \eqref{Iab}, \eqref{Mab} or \eqref{Bpm}.
Let us illustrate, on the simplest $N=2$ example, how an action of the operators \eqref{Mab} creates $(n+1)$-fold degeneracy of the bosonic wave functions
$\tilde{\Omega}^+_{(n_1,l_1)}\otimes\tilde{\Omega}^+_{(n_2,l_2)}$\,, with $n=n_1+n_2$\,.
The action of $M_{1\bar{2}}$ on them is simple:
\bea
   &&M_{1\bar{2}}\left\lbrace\tilde{\Omega}^+_{(n_1,l_1)}\otimes\tilde{\Omega}^+_{(n_2,l_2)}\right\rbrace= \frac{n_2}{2}\left(n_2+\tilde{l}_2-1\right)\left\lbrace\tilde{\Omega}^+_{(n_1+1,l_1)}\otimes\tilde{\Omega}^+_{(n_2-1,l_2)}\right\rbrace,\nn
   &&\left(\mathcal{H}_{\rm conf}-\frac{m\,\tilde{Z}_1}{2}\right)\left\lbrace\tilde{\Omega}^+_{(n_1+1,l_1)}\otimes\tilde{\Omega}^+_{(n_2-1,l_2)}\right\rbrace = m\left(n_1+n_2\right)\left\lbrace\tilde{\Omega}^+_{(n_1+1,l_1)}\otimes\tilde{\Omega}^+_{(n_2-1,l_2)}\right\rbrace.\label{MOmega}
\eea
It just increases the number $n_1$ as  $n_1 \rightarrow n_1 +1$ and decreases $n_2$ as  $n_2 \rightarrow n_2 -1$, so that the total number $n=n_1+n_2$ is not altered.
The action of $M_{2\bar{1}}$ is opposite: $n_1 \rightarrow n_1 -1$, $n_2 \rightarrow n_2 +1$.

A slight modification of the Uhlenbeck tensor \eqref{Iab} by other superconformal SU$(2|1,1)$ generators yields a generalization
of the operator \eqref{inv}, such that it commutes also with the SU$(2|1)$ generators $\tilde{Q}^{i}$, $\bar{\tilde{Q}}_{j}$, $I^{i}_{j}$ and $\tilde{Z}_1$ defined by \eqref{sum}:
\bea
    \tilde{I}_{ab} &=& \frac{1}{m^2}\left[\mathcal{H}_{({\rm conf})\; a}\,\mathcal{H}_{({\rm conf})\; b}-\,\frac{1}{2}\left(T_{a}\bar{T}_{b}+\bar{T}_{a}T_{b}\right)\right]+\frac{1}{4}\,\tilde{Z}_{1a}\tilde{Z}_{1b}-\frac{1}{2}\left(I_a\right)^i_j\left(I_b\right)^j_i\nn
    &&+\,\frac{1}{4m}\left[\left(\tilde{Q}_a\right)^k\left(\bar{\tilde{Q}}_{b}\right)_k-\left(\bar{\tilde{Q}}_{a}\right)_k\left(\tilde{Q}_b\right)^k\right]-\frac{1}{4m}\left[\left(\tilde{S}_a\right)^k\left(\bar{\tilde{S}}_{b}\right)_k-\left(\bar{\tilde{S}}_{a}\right)_k\left(\tilde{S}_b\right)^k\right]\nn
     &&-\,\frac{1}{2m}\left(\mathcal{H}_{({\rm conf})\; a}\,\tilde{Z}_{1b}+\mathcal{H}_{({\rm conf})\; b}\,\tilde{Z}_{1a}\right).
\label{tildeIab}
\eea
In a similar way, the bilinear operator \eqref{Mab} is modified as
\bea
    &&\tilde{M}_{a\bar b}=\frac{1}{2m^2}\,T_{a}\bar{T}_{b}+\frac{1}{4m}\,\left(\tilde{S}_a\right)^i\left(\bar{\tilde{S}}_{b}\right)_i\,.\label{tildeMab}
\eea
Once again, these invariants can be constructed only for new supercharges $\tilde{Q}^i$ and $\bar{\tilde{Q}}_j$\,. No their analogs can be defined for $su(2|1)$
generated by ${Q}^i$ and $\bar{Q}_j$. The algebra of the operators \eqref{tildeIab} and \eqref{tildeMab} is nonlinear and its closure lies in the universal enveloping
of the superconformal algebra $su(2|1,1)$ \eqref{su211m}. The  non-zero commutators of the generators \eqref{tildeIab} and \eqref{tildeMab} are presented in the Appendix B. It is worth
to point out that the crucial property for revealing various degeneracies of the $su(2|1)$ multiplets of the wave functions is the commutativity of $\tilde{M}_{a\bar b}$
and $\tilde{I}_{ab}$ with the SU$(2|1)$ generators $\tilde{Q}^{i}$, $\bar{\tilde{Q}}_{j}$, $I^{i}_{j}$ and $\tilde{Z}_1$ and, hence, with the relevant Casimir operators.  The
precise structure of the closure of the hidden symmetry generators is not too important from this point of view.

\subsection{Products of SU$(2|1)$ representations}

One can consider the degeneracy of eigenvalues of the Casimir operators \eqref{tildeC23} of the $N$ dimensional system, though these eigenvalues
cannot be presented by a generic formula and so each particular $N\geq 2$ model requires a separate analysis.
An additional degeneracy, besides the degeneracy with respect to SU$(2|1)$ generators, comes out with respect
to the hidden symmetry operators \eqref{tildeIab} and \eqref{tildeMab}. Below we present their action as the hidden symmetry operators on the
SU$(2|1)$ multiplets of wave functions. We will always deal with the ``superconformal'' SU$(2|1)$ generated by the generators $\tilde{Q}^i, \tilde{\bar{Q}}_k$
and the relevant SU$(2|1)$ multiplets.

The product of $N$ one dimensional SU$(2|1)$ representations can be decomposed as a non-trivial sum of irreducible SU$(2|1)$ representations \cite{product}.
For simplicity we consider here only $N=2$ case and present the decomposition for the levels $n=0, 1, 2$.

Quantum states are denoted as
\bea
    \left|n , n_1-n_2\,, \tilde{C}_2\,, \tilde{C}_3\right\rangle, \qquad n=n_1+n_2\,,
\eea
where the relevant SU$(2|1)$ Casimir operators are given by the expressions \eqref{tildeC23} with the composite generators \eqref{sum}.
The action of the diagonal elements of \eqref{tildeIab} and \eqref{tildeMab} is defined as
\bea
    \tilde{I}_{11}\left|n , n_1-n_2\,, \tilde{C}_2\,, \tilde{C}_3\right\rangle = \left(\frac{1}{2}-n_1\right)\tilde{l}_1\left|n , n_1-n_2\,, \tilde{C}_2\,, \tilde{C}_3\right\rangle,\nn
    \tilde{I}_{22}\left|n , n_1-n_2\,, \tilde{C}_2\,, \tilde{C}_3\right\rangle = \left(\frac{1}{2}-n_2\right)\tilde{l}_2\left|n , n_1-n_2\,, \tilde{C}_2\,, \tilde{C}_3\right\rangle,\nn
    \tilde{M}_{1\bar 1}\left|n , n_1-n_2\,, \tilde{C}_2\,, \tilde{C}_3\right\rangle = \frac{n_1}{2}\left(n_1+\tilde{l}_1-1\right)\left|n , n_1-n_2\,, \tilde{C}_2\,, \tilde{C}_3\right\rangle,\nn
    \tilde{M}_{2\bar 2}\left|n , n_1-n_2\,, \tilde{C}_2\,, \tilde{C}_3\right\rangle = \frac{n_2}{2}\left(n_2+\tilde{l}_2-1\right)\left|n , n_1-n_2\,, \tilde{C}_2\,, \tilde{C}_3\right\rangle.
\eea
We can also define the action of the operator  $\tilde{I}_{12}+\tilde{M}_{1\bar 2}+\tilde{M}_{2\bar 1}$
\bea
    \left(\tilde{I}_{12}+\tilde{M}_{1\bar 2}+\tilde{M}_{2\bar 1}\right)\left|n , n_1-n_2\,, \tilde{C}_2\,, \tilde{C}_3\right\rangle = n_1\,n_2\,\left|n , n_1-n_2\,, \tilde{C}_2\,, \tilde{C}_3\right\rangle.
\eea
The remaining operators $\tilde{I}_{12}$ and $\tilde{M}_{1\bar 2}-\tilde{M}_{2\bar 1}$ either annihilate a quantum state or change its quantum number $n_1-n_2$\,.

\paragraph{The level $n=0$.} The lowest state with $n=0$ is a product of the single states with $n_1=0$ and $n_2=0$:
\bea
    \left|0\right\rangle \equiv \left|0,0,0,0\right\rangle = \left\lbrace\tilde{\Omega}^+_{(0,l_1)}\otimes\tilde{\Omega}^+_{(0,l_2)}\right\rbrace.
\eea
This singlet state is just the ground state and the operators $\tilde{I}_{12}$, $\tilde{M}_{1\bar 2}$ and $\tilde{M}_{2\bar 1}$ annihilate it.

\paragraph{The level $n=1$.} The level $n=1$ corresponds to a direct sum of two fundamental representations
which are spanned by the states
\bea
    &&\left|1,-1,0,0\right\rangle=\left\lbrace \tilde{\Omega}^+_{(0,l_1)}\otimes\Psi^i_{(1,l_2)}\,,\;
    \tilde{\Omega}^+_{(0,l_1)}\otimes\tilde{\Omega}^+_{(1,l_2)}\right\rbrace,\nn
    &&\left|1,1,0,0\right\rangle=\left\lbrace \Psi^i_{(1,l_1)}\otimes\tilde{\Omega}^+_{(0,l_2)}\,,\;
    \tilde{\Omega}^+_{(1,l_1)}\otimes\tilde{\Omega}^+_{(0,l_2)}\right\rbrace.
\eea
The operator $\tilde{I}_{12}$ mixes these representations according to
\bea
    \tilde{I}_{12}\left|1,-1,0,0\right\rangle = -\,\frac{\tilde{l}_2}{2}\,\left|1,1,0,0\right\rangle,\qquad
    \tilde{I}_{12}\left|1,1,0,0\right\rangle = -\,\frac{\tilde{l}_1}{2}\,\left|1,-1,0,0\right\rangle.
\eea
The operators $\tilde{M}_{1\bar 2}-\tilde{M}_{2\bar 1}$ act as
\bea
    \left(\tilde{M}_{1\bar 2}-\tilde{M}_{2\bar 1}\right)\left|1,-1,0,0\right\rangle = \frac{\tilde{l}_2}{2}\,\left|1,1,0,0\right\rangle,\qquad
    \left(\tilde{M}_{1\bar 2}-\tilde{M}_{2\bar 1}\right)\left|1,1,0,0\right\rangle = -\,\frac{\tilde{l}_1}{2}\,\left|1,-1,0,0\right\rangle.
\eea
Both representations $\left|1,\pm 1,0,0\right\rangle$ are identical to each other with respect to the action of SU$(2|1)$ supersymmetry.
At the same time, they form the simplest non-trivial 2-dimensional representation with respect to the hidden symmetry.

\paragraph{The level $n=2$.} For the excited level $n=2$ there exist three options
\bea
    &&n_1=1,\quad n_2=1,\nn
    &&n_1=0,\quad  n_2=2,\nn
    &&n_1=2,\quad  n_2=0.\label{3opt}
\eea
The product of the fundamental representations with $n_1=1$, $n_2=1$ has the total dimension $3\times 3=9$:
\bea
    \Psi^i_{(1,l_1)}\otimes\Psi^j_{(1,l_2)}\,,\qquad
    \Psi^i_{(1,l_1)}\otimes\tilde{\Omega}^+_{(1,l_2)}\,,\qquad
    \tilde{\Omega}^+_{(1,l_1)}\otimes\Psi^i_{(1,l_2)}\,,\qquad
    \tilde{\Omega}^+_{(1,l_1)}\otimes\tilde{\Omega}^+_{(1,l_2)}\,.
\eea
With respect to the generators \eqref{sum} it splits into a sum of $4$-dimensional typical representation and $5$-dimensional atypical representation.
The atypical representation is spanned by the triplet of bosonic states and the doublet of fermionic states:
\bea
    \left|2,0,0,0\right\rangle = \left\lbrace\Psi^{(i}_{(1,l_1)}\otimes\Psi^{j)}_{(1,l_2)}\,,\;
    \Psi^i_{(1,l_1)}\otimes\tilde{\Omega}^+_{(1,l_2)}-\tilde{\Omega}^+_{(1,l_1)}\otimes\Psi^i_{(1,l_2)}\right\rbrace.
\eea
The operator $\tilde{I}_{12}$ annihilates this state. The typical representation encompasses the states
\bea
    \left|2,0,2,3\right\rangle = \left\lbrace\varepsilon_{ij}\,\Psi^i_{(1,l_1)}\otimes\Psi^j_{(1,l_2)}\,,\;\Psi^i_{(1,l_1)}\otimes\tilde{\Omega}^+_{(1,l_2)}+\tilde{\Omega}^+_{(1,l_1)}\otimes\Psi^i_{(1,l_2)}\,,\;\tilde{\Omega}^+_{(1,l_1)}\otimes\tilde{\Omega}^+_{(1,l_2)}\right\rbrace.
    \label{4typ}
\eea

The remaining two options associated with \eqref{3opt} are given by the states
\bea
    &&\left|2,-2,2,3\right\rangle=\left\lbrace\tilde{\Omega}^+_{(0,l_1)}\otimes\tilde{\Omega}^-_{(2,l_2)}\,,\;
    \tilde{\Omega}^+_{(0,l_1)}\otimes\Psi^i_{(2,l_2)}\,,\;
    \tilde{\Omega}^+_{(0,l_1)}\otimes\tilde{\Omega}^+_{(2,l_2)}\right\rbrace,\nn
    &&\left|2,2,2,3\right\rangle=\left\lbrace \tilde{\Omega}^-_{(2,l_1)}\otimes\tilde{\Omega}^+_{(0,l_2)}\,,\;
    \Psi^i_{(2,l_1)}\otimes\tilde{\Omega}^+_{(0,l_2)}\,,\;
    \tilde{\Omega}^+_{(2,l_1)}\otimes\tilde{\Omega}^+_{(0,l_2)}\right\rbrace.
\eea
So we face three identical SU$(2|1)$ representations $\left|2,0,2,3\right\rangle$, $\left|2,-2,2,3\right\rangle$
and $\left|2,2,2,3\right\rangle$ on the level $n=2$, since they possess the same values of the SU$(2|1)$ Casimir operators
\bea
    \tilde{C}_2 = 2,\qquad \tilde{C}_3 = 3.
\eea
However, the operators $\tilde{I}_{bb}\tilde{Z}_{1a}-\tilde{I}_{aa}\tilde{Z}_{1b}=\tilde{l}_a\tilde{l}_b\left(n_a-n_b\right)$ take different values on these SU$(2|1)$
multiplets and so discriminate them. The operators $\tilde{I}_{12}$\,, $\tilde{M}_{1\bar 2}-\tilde{M}_{2\bar 1}$ mix them among one another.

The further consideration of irreducible SU$(2|1)$ representations for higher levels $n>2$ brings us
to the same conclusions about the action of the operators $\tilde{I}_{ab}$ and $\tilde{M}_{a\bar b}$\,. The values of the diagonal operators
$\tilde{I}_{aa}$ uniquely mark each representation in the set of identical ({\it i.e.}, having the same values of the Casimir operators)
SU$(2|1)$ irreducible representations, while the off-diagonal operators mix these representations among themselves
as the appropriate packages of bosonic and fermionic states.

\section{Summary and outlook}
In this paper we studied the quantum mechanics of SU$(2|1)$ supersymmetric extension of
 the S.-W. system on the complex Euclidian space  $\mathbb{C}^N$ interacting with an external constant magnetic field \cite{Shmavonyan}.
 This supersymmetric system  can be considered as a unification of $N$ non-interacting $\mathbb{C}^1$ S.-W. systems.
Accordingly, we first quantized the model on $\mathbb{C}^1$ and then generalized the consideration to the case of $\mathbb{C}^N$.
We constructed the complete space of the wave functions and found the relevant energy spectrum. We studied how
 all bosonic and fermionic states are distributed over the irreducible representations of the supergroup SU$(2|1)$.
 We also showed that the bosonic S.-W. model possesses
 conformal symmetry SO$(2,1)$ (see Appendix C). In the supersymmetric case we redefined the Hamiltonian as \eqref{redef} and showed that
 it exhibits SU$(2|1,1)$ superconformal symmetry which serves as the spectrum-generating symmetry on the full set of the quantum states.

The wave functions of supersymmetric quantum S.-W. system on $\mathbb{C}^N$ were constructed as products of
$N$ wave functions  of the $\mathbb{C}^1$ models. Correspondingly, on these products irreducible SU$(2|1)$ representations are realized.
For simplicity we considered the case of $N=2$ that already amounts to a non-trivial sum of irreducible SU$(2|1)$ representations.
Also, the generalization to $\mathbb{C}^N$ reveals hidden symmetry generators \eqref{tildeIab} which correspond to
a supersymmetrization of Uhlenbeck tensor \cite{Shmavonyan,ClSW}. It is responsible for the degeneracy of
the wave functions belonging to irreducible SU$(2|1)$ representations. The general expressions for the hidden symmetry generators in
terms of products of the generators of the superconformal algebra $su(2|1,1)$ were found.

It would be interesting to consider, along the same lines, quantum deformed SU$(2|1)$ extensions of other K\"ahler oscillator models, e.g., of the
$\mathbb{CP}^N$ one. These models are not superconformal, so their quantum analysis should be similar to what has been performed in Section 2.
On the other hand, a non-trivial multi-particle extension of the  SU$(2|1)$ $\mathbb{C}^1$ S.-W. model could be a complex $N$-particle interacting system
of the Calogero-Moser type, hopefully preserving the superconformal invariance of the one-particle $\mathbb{C}^1$ model. Then the whole consideration
of Sections 3 and 4 based on the superconformal group SU$(2|1,1)$ could be applicable.

Finally, let us notice that the quantum $\mathbb{C}^1$ S.-W. system
without magnetic field
(also known as a ``circular oscillator with ring-shaped potential'')  was used in a more phenomenological setting for the study of the particle behavior
in the two-dimensional quantum ring \cite{ring}.
Respectively, the $\mathbb{C}^N$ S.-W. system  with coincident parameters $g_a$
can be interpreted as an ensemble of $N$ {\sl free} particles in a single quantum ring interacting with a constant magnetic field orthogonal to the plane.
It would be interesting to reveal possible physical implications of the quantum SU(2$|1)$ supersymmetric version of this system within such an interpretation.
\section*{Acknowledgments}
The authors thank Hovhannes Shmavonyan for interest in the work. A.N. thanks Ruben Mkrtchyan for useful comments. The work of A.N. was supported by the Armenian Science Committee and Russian Foundation for Basic Research within the joint research project 20RF-023. E.I. and S.S. acknowledge support from the RFBR grant No. 20-52-05008 Arm-a and a grant of the Ter-Antonyan--Smorodinsky Program. Their research was also supported in part by the Ministry of Science and Higher Education of Russian Federation, project FEWF-2020-0003.

\appendix

\section{More on the integrals of motion \eqref{Aa}}
The fermionic integrals of motion $A_a$  can further be expressed via other, odd  integrals of motion, as follows
\bea
    A_a-\frac{I_a^2}{3}&=&\frac{1}{8m}\left[\left(\Xi_{a}\right)^k\left(\bar{\mathcal{S}}_{a}\right)_k-\left(\bar{\Xi}_{a}\right)_k\left(\mathcal{S}_{a}\right)^k-\frac{m}{2}\,\left(\bar{\Xi}_{a}\right)_k\left(\Pi_{a}\right)^k\right]\nn
    &&-\,\frac{1}{8m}\left[\left(\Pi_{a}\right)^k\left(\bar{\mathcal{Q}}_{a}\right)_k-\left(\bar{\Pi}_{a}\right)_k\left(\mathcal{Q}_{a}\right)^k+\frac{m}{2}\,\left(\bar{\Pi}_{a}\right)_k\left(\Xi_{a}\right)^k\right],
\eea
where
\bea
&&\left(\mathcal{Q}_{a}\right)^i:= \left(\tilde{Q}_{a}\right)^i\left(T_a\right)^{1/2},\qquad \left(\bar{\mathcal{Q}}_{a}\right)_j=\left(\bar{T}_a\right)^{1/2}\left(\bar{\tilde{Q}}_{a}\right)_j\,,\nn
&&\left(\mathcal{S}_{a}\right)^i:=\left(\tilde{S}_{a}\right)^i\left(\bar{T}_a\right)^{1/2},\qquad \left(\bar{\mathcal{S}}_{a}\right)_j=\left(T_a\right)^{1/2}\left(\bar{\tilde{S}}_{a}\right)_j\,,\nn
&&\left(\Pi_{a}\right)^i:=\left(\tilde{Q}_{a}\right)^i\left(\bar{T}_a\right)^{-1/2},\qquad\left(\bar{\Pi}_{a}\right)_j:=\left(T_a\right)^{-1/2}\left(\bar{\tilde{Q}}_{a}\right)_j\,,\nn
&&\left(\Xi_{a}\right)^i:=\left(\tilde{S}_{a}\right)^i\left(T_a\right)^{-1/2},\qquad\left(\bar{\Xi}_{a}\right)_j:=\left(\bar{T}_a\right)^{-1/2}\left(\bar{\tilde{S}}_{a}\right)_j\,.
\eea
Using the $su(2|1,1)$ (anti)commutation relations \eqref{su211m}, it is straightforward to check that all these fermionic operators
commute with the Hamiltonian $\mathcal{H}_{({\rm conf})\; a}$\,. Note that the generators $T_a, \bar{T}_a$ in the present model
start with non-zero numerical constants (as follows from the definitions \eqref{HKD} and \eqref{confH0}, \eqref{DefDK}), so the operators invert to them,
equally as their square roots, are well defined.
In fact, not all of these odd integrals of motion are independent. We can choose among them the basis $\left(\mathcal{Q}_{a}\right)^i$, $\left(\bar{\mathcal{Q}}_{a}\right)_j$
and represent the remaining ones, taking into account the relations \eqref{su211m}, as
\bea
    &&\left(\mathcal{S}_{a}\right)^i=\frac{2}{m}\left[\left(\mathcal{Q}_{a}\right)^i
    -\left(T_a\right)^{1/2}\left(\bar{T}_a\right)^{1/2}\left(\mathcal{Q}_{a}\right)^i\left(T_a\right)^{-1/2}\left(\bar{T}_a\right)^{-1/2}
   \right]\left(T_a\right)^{1/2}\left(\bar{T}_a\right)^{1/2},\nn
    &&\left(\bar{\mathcal{S}}_{a}\right)_j=\frac{2}{m} \left(T_a\right)^{1/2}\left(\bar{T}_a\right)^{1/2}\left[\left(\bar{\mathcal{Q}}_{a}\right)_j
    -\left(T_a\right)^{-1/2}\left(\bar{T}_a\right)^{-1/2}\left(\bar{\mathcal{Q}}_{a}\right)_j\left(T_a\right)^{1/2}\left(\bar{T}_a\right)^{1/2}\right],\nn
    &&\left(\Pi_a\right)^i=\left(\mathcal{Q}_{a}\right)^i\left(T_a\right)^{-1/2}\left(\bar{T}_a\right)^{-1/2},\nn
    &&\left(\bar{\Pi}_{a}\right)_j=\left(T_a\right)^{-1/2}\left(\bar{T}_a\right)^{-1/2}\left(\bar{\mathcal{Q}}_{a}\right)_j\,,\nn
    &&\left(\Xi_{a}\right)^i=\frac{2}{m}\left[\left(\mathcal{Q}_{a}\right)^i-\left(T_a\right)^{1/2}\left(\bar{T}_a\right)^{1/2}\left(\mathcal{Q}_{a}\right)^i\left(T_a\right)^{-1/2}\left(\bar{T}_a\right)^{-1/2}\right],\nn
    &&\left(\bar{\Xi}_{a}\right)_j=\frac{2}{m}\left[\left(\bar{\mathcal{Q}}_{a}\right)_j-\left(T_a\right)^{-1/2}\left(\bar{T}_a\right)^{-1/2}\left(\bar{\mathcal{Q}}_{a}\right)_j\left(T_a\right)^{1/2}\left(\bar{T}_a\right)^{1/2}\right].
\eea
The integrals of motion $\left(T_a\right)^{\pm 1/2}\left(\bar{T}_a\right)^{\pm 1/2}$ can be argued to functionally depend on $M_{a\bar{a}}$ and $\mathcal{H}_{({\rm conf})\; a}$\,.

Hence, our system has $2N$ commuting bosonic integrals $\mathcal{H}_{({\rm conf})\; a}$\,, $L_a$ and $4N$ Hermitian (or $2N$ complex)
fermionic ones $\left(\mathcal{Q}_{a}\right)^i$, $\left(\bar{\mathcal{Q}}_{a}\right)_j$\,,
{\it i.e.} it is integrable in the sense of supergeneralization of Liouville theorem \footnote{The Hamiltonian system on $(2k|n)$-dimensional symplectic supermanifold is integrable
if it possesses $n$ functionally independent  fermionic odd integrals and $k$ functionally independent {\sl commuting} bosonic integrals, see \cite{shander}.}.

\section{Commutators of super Uhlenbeck tensor}
The commutator of off-diagonal elements ($a\neq b,\; b\neq c,\; c\neq a$) of the hidden symmetry generators $\tilde{I}_{ab}$ introduced in \eqref{tildeIab} is given by
\bea
    \left[\tilde{I}_{ab},\tilde{I}_{bc}\right]&=&\tilde{T}_{abc}+\frac{\tilde{Z}_{1b}}{2}\left(\tilde{M}_{a\bar c}-\tilde{M}_{c\bar a}\right),\label{OFF}
\eea
where the totally antisymmetric tensor $\tilde{T}_{abc}$ is defined as
\bea
    \tilde{T}_{abc}&=&\frac{1}{12}\left(I_a\right)^i_j\left[\left(I_b\right)^j_k\left(I_c\right)^k_i-\left(I_b\right)^k_i\left(I_c\right)^j_k\right]+\frac{\mathcal{H}_{({\rm conf})\; a}}{2m^3}\left(\bar{T}_{b}T_{c}-T_{b}\bar{T}_{c}\right)-\frac{\tilde{Z}_{1a}}{4m^2}\left(\bar{T}_{b}T_{c}-T_{b}\bar{T}_{c}\right)\nn
    &&+\,\frac{\tilde{Z}_{1a}}{16m}\left\lbrace\left[\left(\tilde{S}_b\right)^k\left(\bar{\tilde{S}}_{c}\right)_k+\left(\bar{\tilde{S}}_{b}\right)_k\left(\tilde{S}_c\right)^k\right]+\left[\left(\tilde{Q}_b\right)^k\left(\bar{\tilde{Q}}_{c}\right)_k+\left(\bar{\tilde{Q}}_{b}\right)_k\left(\tilde{Q}_c\right)^k\right]\right\rbrace\nn
    &&-\,\frac{\mathcal{H}_{({\rm conf})\; a}}{8m^2}
    \left\lbrace\left[\left(\tilde{S}_b\right)^k\left(\bar{\tilde{S}}_{c}\right)_k+\left(\bar{\tilde{S}}_{b}\right)_k\left(\tilde{S}_c\right)^k\right]
    +\left[\left(\tilde{Q}_b\right)^k\left(\bar{\tilde{Q}}_{c}\right)_k+\left(\bar{\tilde{Q}}_{b}\right)_k\left(\tilde{Q}_c\right)^k\right]\right\rbrace\nonumber \\
    &&+\,\frac{\left(I_a\right)^j_i}{8m}\left\lbrace\left[\left(\tilde{S}_b\right)^i\left(\bar{\tilde{S}}_{c}\right)_j
    +\left(\bar{\tilde{S}}_{b}\right)_j\left(\tilde{S}_c\right)^i\right]-\left[\left(\tilde{Q}_b\right)^i\left(\bar{\tilde{Q}}_{c}\right)_j
    +\left(\bar{\tilde{Q}}_{b}\right)_j\left(\tilde{Q}_c\right)^i\right]\right\rbrace\nn
    &&+\,\frac{T_a}{8m^2}\left[\left(\tilde{Q}_b\right)^k\left(\bar{\tilde{S}}_{c}\right)_k+\left(\bar{\tilde{S}}_{b}\right)_k\left(\tilde{Q}_c\right)^k\right]
    +\frac{\bar{T}_a}{8m^2}\left[\left(\tilde{S}_b\right)^k\left(\bar{\tilde{Q}}_{c}\right)_k+\left(\bar{\tilde{Q}}_{b}\right)_k\left(\tilde{S}_c\right)^k\right]\nn
    &&+ \,\left({\rm cyclic\;permutation\;of}\;\;abc\right).
\eea
The rest of the non-vanishing commutators of the algebra is given by
\bea
    \left[\tilde{I}_{aa},\tilde{I}_{ab}\right]&=&\tilde{Z}_{1a}\left(\tilde{M}_{a\bar b}-\tilde{M}_{b\bar a}\right).
\eea

\section{Conformal symmetry in the bosonic sector}

The $\mathbb{C}^N$ S.-W. quantum system can be conveniently considered in the framework of conformal quantum mechanics.
This peculiar feature was not noticed in \cite{Shmavonyan}.

Let us concentrate on the $\mathbb{C}^1$ case. The Hamiltonian in the presence of magnetic field $B$ reads
\bea
    H_{\rm SW} = -\,\partial_{\bar{z}}\partial_{z}+\frac{B}{2}\,L
    + \frac{m^2z\bar{z}}{4}+\frac{g^2}{4z\bar{z}}\,, \quad L = z\partial_{z}-\bar{z}\partial_{\bar{z}} \,.\label{BosSW}
\eea
The spectrum of $H_{\rm SW}$ is given by
\bea
    H_{\rm SW}\,\Phi_{(n,l)} = \left[m\left(n+\frac{\tilde{l}}{2}+\frac{1}{2}\right)+\frac{B\,l}{2}\right]\Phi_{(n,l)}\,, \quad L\,\Phi_{(n,l)} = l\Phi_{(n,l)}\,,
\eea
where
\bea
    \Phi_{(n , l)} = n!\,\left(z\bar{z}\right)^{\frac{\tilde{l}}{2}}\left(z/\bar{z}\right)^{\frac{l}{2}}
    e^{-\frac{mz\bar{z}}{2}}L^{(\tilde{l}\,)}_{n}\left(mz\bar{z}\right),\qquad
    \tilde{l} = \sqrt{l^2 + g^2}\,.
\eea
One sees that the presence of magnetic field $B$ affects the energy spectrum.

On the other hand, we can represent the Hamiltonian \eqref{BosSW} in the form
\bea
    H_{\rm SW}=\mathcal{H}_{\rm conf}+\frac{B}{2}\,L\,,\label{magneto}
\eea
where $\mathcal{H}_{\rm conf}$ is a conformal Hamiltonian of the trigonometric type,
\bea
    \mathcal{H}_{\rm conf}=-\,\partial_{\bar{z}}\partial_{z} + \frac{m^2z\bar{z}}{4}+\frac{g^2}{4z\bar{z}}\,.
\eea
One can check that $L$ commutes with the conformal generators $T$, $\bar{T}$ and $\mathcal{H}_{\rm conf}$ satisfying the algebra \eqref{so21}, where
\bea
    T=\mathcal{H}_{\rm conf}- \frac{m^2z\bar{z}}{2} +\frac{m}{2}\left(z\partial_z+\bar z\partial_{\bar z}+1\right),\quad
    {\bar T}=\mathcal{H}_{\rm conf}- \frac{m^2z\bar{z}}{2} -\frac{m}{2}\left(z\partial_z+\bar z\partial_{\bar z}+1\right).
\eea
The introduction of magnetic field according to \eqref{magneto} modifies the conformal algebra $so(2,1)$ written in terms of the original Hamiltonian $H_{\rm SW}$ as
\bea
    \left[\bar{T},T\right]=2m\,H_{\rm SW} - B\,L\,,\quad\left[H_{\rm SW} \,, \bar{T}\right]=-\,m\,\bar{T},\quad \left[H_{\rm SW} \,, T\right]=m\,T,
\eea
with $L$ playing the role of a central charge. The conformal algebra has the standard form \eqref{so21} just in the basis with the conformal Hamiltonian,
with $L$ becoming an external generator commuting with all conformal generators.

The generators $T$, $\bar{T}$ and $\mathcal{H}_{\rm conf}$ act on the wave functions $\Phi_{(n,l)}$,
as
\bea
    &&T\,\Phi_{(n,l)}=m\,\Phi_{(n+1,l)}\,,\qquad \bar{T}\,\Phi_{(n,l)}=n\left(n+\tilde{l}\right)m\,\Phi_{(n-1,l)}\,,\nn
    &&\mathcal{H}_{\rm conf}\,\Phi_{(n,l)} = m\left(n+\frac{\tilde{l}}{2}+\frac{1}{2}\right)\Phi_{(n,l)}\,.
\eea
We observe that the explicit dependence on the external magnetic field disappears in the spectrum of the conformal Hamiltonian $\mathcal{H}_{\rm conf}$\,.

The relevant Casimir operator of $so(2,1)$ defines an irreducible representation of the tower of states $\Phi_{(n,l)}$ for $n=0, 1, 2, \ldots$ and a fixed $l$:
\bea
    \frac{1}{m^2}\left[\mathcal{H}_{{\rm conf}}\,\mathcal{H}_{{\rm conf}}-\frac{1}{2}\left(T\bar{T}+\bar{T}T\right)\right]\Phi_{(n,l)}=
    \frac{\left(\tilde{l}^2-1\right)}{4}\,\Phi_{(n,l)}\,.\label{ConfCas}
\eea
So, at each $l$ the quantum states constitute an infinite-dimensional irreducible representation of $so(2,1) \sim sl(2,R)$. Hence the conformal
algebra serves as the spectrum-generating algebra of the quantum $\mathbb{C}^1$ S.-W. model. Note that the relation \eqref{ConfCas} immediately
follows from the operator identity
\bea
\mathcal{H}_{{\rm conf}}\,\mathcal{H}_{{\rm conf}}-\frac{1}{2}\left(T\bar{T}+\bar{T}T\right) = \frac{m^2}{4}\left(L^2 + g^2 -1\right),\label{B10}
\eea
which holds upon substitution of the explicit expressions for the involved generators.

Passing to the $\mathbb{C}^N$ case comes about quite analogously to the full supersymmetric case, with the hidden symmetry generators defined
by eqs. \eqref{Iab0} -- \eqref{MDef}. In particular, for $N\,{=}\,2$,
by analogy with \eqref{MOmega}, the action of the operators \eqref{Mab} accounts for $(n+1)$-fold degeneracy of the bosonic wave functions
$\Phi_{(n_1,l_1)}\otimes\Phi_{(n-n_1,l_2)}$\,, where $n_1=0, 1, 2, \ldots n$\,.

In the general case of $\mathbb{C}^N$ S.-W. quantum system, the degeneracy with respect to \eqref{Mab} is given by the binomial coefficient $C^n_{n+N} = \frac{(n+N)!}{n!N!}$\,.

\end{document}